\documentclass{elsart}
\usepackage{graphicx}
\usepackage{amssymb}

\begin{document}

\hspace{6cm} REPORT NPI ASCR \v{R}E\v{Z}, EXP-01/2009 \\
\\
\begin{frontmatter}
\title{Long term stability of the energy of conversion electrons
emitted from solid $^{83}$Rb/$^{83\rm{m}}$Kr source}
\author[NPI]{D.~V\'{e}nos\corauthref{co}},
\corauth[co]{Tel.: +420 212 241 677; fax:
             +420 220 941 130}
\ead{venos@ujf.cas.cz}
\author[NPI]{J.~Ka\v{s}par},
\author[WWU,NPI]{M.~Zbo\v{r}il},
\author[NPI]{O.~Dragoun},
\author[JGU]{J.~Bonn},
\author[JINR]{A.~Koval\'ik},
\author[NPI]{O.~Lebeda},
\author[NPI]{M.~Ry\v{s}av\'y},
\author[KFZ]{K.~Schl\"osser},
\author[NPI]{A.~\v{S}palek},
\author[WWU]{Ch.Weinheimer}
\address[NPI]{Nuclear Physics Institute,
              Academy of Sciences of the Czech Republic,
              CZ-250 68 \v Re\v z near Prague,
              Czech Republic}
\address[WWU]{Institut f\"{u}r Kernphysik, University of M\"{u}nster,
              D-48149 M\"{u}nster, Germany}
\address[JGU]{Institut f\"{u}r Physik, University of Mainz,
              D-55099 Mainz, Germany}
\address[KFZ]{Forschungszentrum Karlsruhe, D-76344 Eggenstein-Leopoldshafen,
              Germany}
\address[JINR]{Joint Institute of Nuclear Research, Dubna, Russia}
\begin{abstract}
The mono-energetic conversion electrons from the decay of
$^{83\rm{m}}$Kr represent a unique tool for the energy
calibration, energy scale monitoring and systematic studies of the
tritium beta spectrum measurement in the neutrino mass experiment
KATRIN. For this reason, the long term stability of energy of the
7.5~keV and 17.8~keV conversion electrons populated in the decay
of solid $^{83}$Rb/$^{83\rm{m}}$Kr vacuum evaporated sources was
examined by means of two electron spectrometers.
\end{abstract}
\end{frontmatter}

\section{Introduction}
\label{1intro} The knowledge of the absolute value of neutrino
mass is crucial for understanding the fermion masses. Moreover,
massive neutrinos and their properties play an important role in
astrophysics and cosmology for explaining the hot dark matter.
Current upper limits ascribed to the effective neutrino mass,
$m_{\nu_{e}}^{(eff)}$,~2~eV/$c^2$ (95\% C.L.) published by the
particle data group \cite{Yao06} is based on the limits from the
tritium $\beta$-decay experiments performed in Mainz
\cite{Kra05} --- 2.3~eV/$c^2$ and in Troitsk
\cite{Lob03} --- 2.1~eV/$c^2$. With respect to the great importance
of knowledge of the neutrino mass a next generation tritium
$\beta$-decay experiment KATRIN (KArlsruhe TRItium Neutrino
experiment) is being built \cite{Ang05}. The aim of the project is
to determine $m_{\nu_{e}}^{(eff)}$ or its upper limit with the
0.2~eV/$c^2$ (90\% C.L.) sensitivity after full three years of
data taking.

The KATRIN experimental setup consists of several components:
windowless gaseous tritium source with differential pumping and
cryosorption sections, pre- and main electrostatic retarding
spectrometers, detector for $\beta$ counting and monitoring
electron spectrometer. The spectrometers of MAC-E-Filter (Magnetic
Adiabatic Collimation Electrostatic Filter) type will work as
electrostatic filters --- all electrons with the energy high enough
to overcome the electric potential set can be detected. The
pre-spectrometer with suitably pre-set electric potential will
prevent huge amount of low energy electrons from tritium decay
(carrying no information about the neutrino mass) to enter the
main spectrometer and thus causing undesirable background. As a
primary result, the integral beta spectrum measured by main
spectrometer will be recorded. The narrow range of (-30,+5)~eV
around the 18575~eV endpoint of the tritium $\beta$-spectrum will
be inspected in detail as it is most sensitive region to neutrino
mass \cite{Ang05}. The reliable determination of the neutrino mass
squared from fit of the theoretical $\beta$-spectrum
to measured one does not require a precise knowledge of the
absolute retarding voltage applied to the main spectrometer.
However, the numerical calculations have shown that an
unrecognized smearing and/or a shift of the retarding voltage by $\pm
50$~mV would result in systematic error of the fitted neutrino
mass by $\pm 0.04$~eV which is a non-negligible part of the
expected KATRIN sensitivity of 0.2~eV \cite{Ang05,Kas04}.
The energy spectrum is scanned by varying low voltage applied to
the tritium source. The low voltage can be precisely measured by
modern voltmeters.   The long term monitoring of the high
retarding potential of 18.6~kV at a level of about $\pm 60$~mV in 2~months
($\pm 3$~ppm in 2~months) represents a challenge. For
this reason the KATRIN retarding voltage will be measured by two
methods. As the first method the standard procedure will be
applied --- a high-precision voltmeter together with a high voltage
resistance divider, directly attached to the main spectrometer
retarding potential \cite{Thu07}. The second method will use a monitoring
spectrometer principally of the same type as KATRIN main
spectrometer - upgraded Mainz neutrino mass spectrometer
MAC-E-Filter. The retarding voltage of both the monitoring and main
spectrometers will be common (from the same power supply). The
monitoring spectrometer will be equipped with a source of
mono-energetic electrons the energy of which will be stable and
close to the tritium endpoint energy. The measurement of the
spectrum will be provided by means of scanning low voltage
power supply attached to the source itself. Any change of the
energy of mono-energetic electrons measured on the monitoring
spectrometer will indicate a possible change of the common
retarding voltage.

The conversion electrons of the 9.4 and 32.2~keV transitions of
the isomeric state of $^{83\rm{m}}$Kr proved to be useful for the
calibration and systematic studies in previous tritium neutrino
experiments as described in \cite{Rob91,Sto95,Pic9a,Ase00,Bor03}.
Particularly, the energy of K conversion electrons of the 32~keV
transition for free krypton atom (denoted as K-32) of
$17824.3(5)$~eV \cite{Ven06} is rather close to the tritium
endpoint. In those tritium experiments gaseous and condensed
$^{83\rm{m}}$Kr sources were utilized. The new condensed source
had to be prepared every few hours due to 1.83~h half life of
$^{83\rm{m}}$Kr. To overcome this, a not easy technique of
continuously condensed krypton source was suggested and tested
\cite{Ost08}. From this point of view a solid source
$^{83}$Rb/$^{83\rm{m}}$Kr would be advantageous --- its lifetime
is governed by the half life of $^{83}$Rb amounting to 86.2~d.
It seems that the $^{83}$Rb/$^{83\rm{m}}$Kr source could be used
as the calibration source of KATRIN in the monitoring spectrometer.
On the other side, the more complex condensed $^{83\rm{m}}$Kr
source, which was proven to be able to provide long-term stability,
will be used for less-frequent calibrations of the KATRIN main
spectrometer.

The vacuum evaporated solid $^{83}$Rb/$^{83\rm{m}}$Kr sources on
aluminium backing were successfully used in Auger and conversion
electron measurements with the electrostatic spectrometer in Dubna
\cite{Kov92,Kov93}. These sources are physico-chemically stable
both in vacuum and air for several months. The energy of the K-32
line was reported to be stable within experimental error of 1.1~eV
\cite{Kov01}. Although $^{83\rm{m}}$Kr is a gas and can have
limited retention in a solid radioactive source, the effect of
limited krypton retention was not reported in the above cited
works. Regardless many years effort of nuclear electron
spectroscopists there are no solid radioactive sources of the
conversion electron useable for the energy calibration at
sub-electronvolt level. The reason is a high sensitivity of the
electron binding energies to the physico-chemical environment of
the atoms of decaying nuclei. (On the other side this feature,
exhibited also by the innermost atomic shells, is basis of the
powerful X-ray photoelectron spectroscopy applied to surface
analysis, see e.g. \cite{Bri03}). The necessary condition for
precise determination of the electron energy is an instrumental
resolution of a few~eV that allows to detect so called zero-energy
loss peak in the measured spectrum. This peak corresponds to
conversion electrons that leave the radioactive source without any
energy loss. The width of the zero-energy loss peak is determined
by the natural widths of the corresponding atomic subshell and
nuclear state and the width of the instrumental response function.
A vacuum evaporation of the carrier free activity on the flat
source backing was found to be one of the most promising methods
for preparation of these radioactive sources. Using the practice
of \cite{Kov93} the technique of purification of rubidium activity
in a micro-chromatography column with subsequent vacuum
evaporation is available. We have to our disposal two electron
spectrometers, one at NPI \v{R}e\v{z} ESA12 (differential
spectrometer, energy range of 0--20~keV with moderate resolution
$\Delta E/ E = 0.011$ in the basic measurement mode and resolution
$\Delta E$ of 1 and 3~eV for energies 2 and 7~keV in the
retardation mode with reduced luminosity) and the second one at
Mainz University (integral spectrometer, energy resolution  $E /
\Delta E=10000 \div 20000$ for the K-32 electrons, energy range
7-35~keV).

Previously a good experience was obtained as for the reproducibility
of the energy of the $^{99\rm{m}}$Tc M$_{5}$-2.2 conversion line
with energy of 1917 eV --- at level of about 0.040~eV \cite{Fis96}.
For this result the sources were periodically, once a day,
prepared by electrolysis onto platinum substrate and immediately
measured. On the other side, a change of energy of the K-14
$^{57}$Co conversion line after storage (at air) of a source
prepared on aluminium backing was observed.  Also a dependence of
the K-14 line energy on backing material was noticed \cite{Spa02}.

For the reasons indicated above the development of the solid
$^{83}$Rb/$^{83\rm{m}}$Kr conversion electrons source for
calibration and monitoring purposes at sub-electronvolt level was started.

In this work we report about the experimental methods,
instrumentation equipments, conversion electron sources
production, measurements of the energy of the conversion
electrons and about first results.

\section{Experimental method}
\label{2exper}
\subsection{Conversion electrons from $^{83}$Rb decay}
\label{21conv} 
\begin{figure}[th]
\includegraphics[100pt,333pt][419pt,510pt]{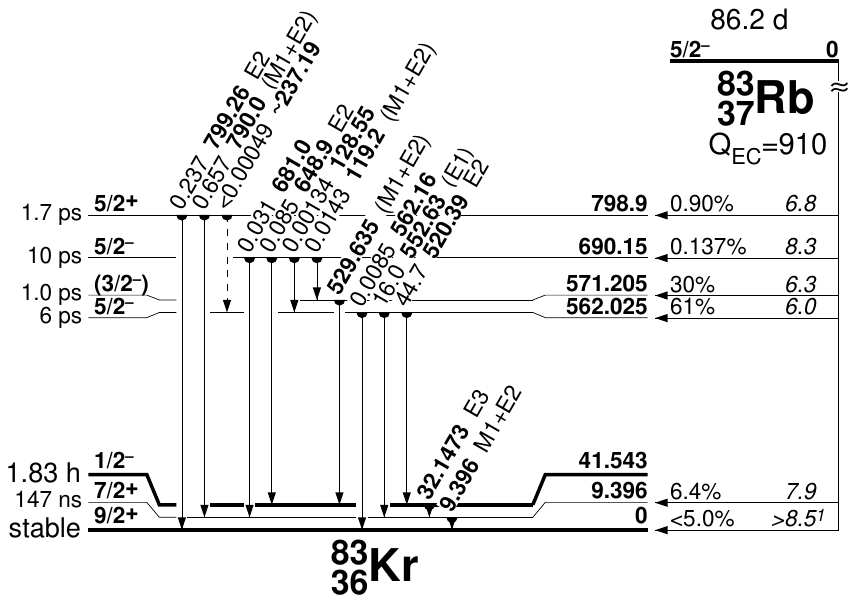}
\caption[fig1]{Decay scheme of $^{83}$Rb taken from \cite{Fir98}.}
\label{schemea}
\end{figure}

\begin{figure}[h]
 \begin{center}
 \includegraphics[bb=19pt 345pt 449pt 644pt, width=12cm]{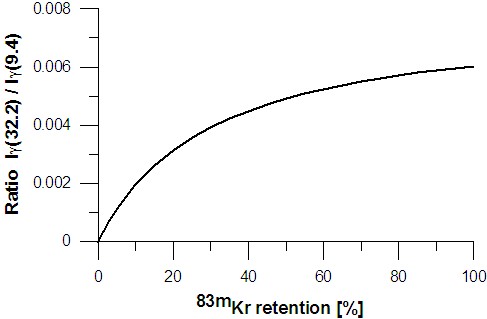}
 \caption[fig2]{ The dependence of the 32.2 and 9.4 $\gamma$-ray
  intensity ratio on the retention of
  $^{83\rm{m}}$Kr in $^{83}$Rb/$^{83\rm{m}}$Kr source.}
 \label{retention}
 \end{center}
\end{figure}

The $^{83}$Rb nuclide decays by electron capture
with a half-life of T$_{1/2}$=86.2(1)~d to form the short lived
isomeric state $^{83\rm{m}}$Kr with T$_{1/2}$=1.83(2)~h, i.e. the
$^{83\rm{m}}$Kr isomer is continuously generated by the decay of
$^{83}$Rb, see the decay scheme in Fig.~\ref{schemea}.
The conversion electrons produced in electromagnetic transitions
from the krypton nuclear levels above the isomeric state are not
usable for systematic studies and calibrations for the tritium
neutrino mass experiment as their energies are too high.
Conversely, the isomeric state decays via a cascade of suitable
low energy transitions of 32.2 and 9.4~keV possessing high
intensity of conversion electrons. The first transition with E3
multipolarity is especially highly converted --- the total
conversion coefficient calculated using the tables \cite{Ros78}
amounts to $\alpha_{tot}=I_{K+L+..}/I_{\gamma} = 2010$. The second
transition of 9.4~keV energy is practically a pure M1
transition --- the amount of E2 multipolarity admixture is given by
the mixing parameter $\delta$=0.0130(8). The energies and
intensities of the $\gamma$-transitions and the related conversion
electrons (with intensity larger than 0.1~\%) from $^{83}$Rb decay
are listed in Table~\ref{tab1}.
\begin{table}[h]
 \begin{center}
 \caption[tabl2] {The $\gamma$-radiations and intensive conversion electrons
  for $^{83\rm{m}}$Kr transitions of 9.4 and 32.2~keV}
 \begin{tabular}{ccccc}
 \hline
Energy$^{a}$   &  Transition & Intensity$^{b}$ &  Line width$^{c}$ & Line width$^{d}$ \\
  eV     &          type     & \% per $^{83}$Rb decay\ & eV & eV \\
 \hline
9405.9(8)\cite{Pic9a}  & $\gamma$  & 6.1(14)\cite{Vai76}&  -       \\
7481.2(11)             & L$_{1}$   & 73.2               &  5.3(4)  & 3.75\\
7674.9(9)              & L$_{2}$   & 8.24               &  1.84(5) & 1.25\\
7727.4(9)              & L$_{3}$   & 5.28               &  1.40(2) & 1.19\\
9113.0(9)              & M$_{1}$   & 12.1               &  4.27(5) & 3.5\\
9183.6(8)              & M$_{2}$   & 1.35               &  1.99(32)& 1.6\\
9191.4(8)              & M$_{3}$   & 1.03               &  1.66(8) & 1.1\\
9378.4(8)              & N$_{1}$   & 1.49               &  0.19(4) & 0.4\\
9391.7(8)              & N$_{2}$   & 0.12               &  0.59(4) &  - \\
32151.7(5)\cite{Ven06} & $\gamma$  & 0.037(5)\cite{Vai76}& -       \\
17824.3(5)$^{e}$       & K         & 17.6               &  2.83(12)& 2.71\\
30226.9(9)             & L$_{1}$   & 1.17               &  5.3(4)  & 3.75\\
30420.6(7)             & L$_{2}$   & 18.2               &  1.84(5) & 1.25\\
30473.1(7)             & L$_{3}$   & 28.3               &  1.40(2) & 1.19\\
31858.7(6)             & M$_{1}$   & 0.19               &  4.27(5) & 3.5\\
31929.3(5)             & M$_{2}$   & 3.10               &  1.99(32)& 1.6\\
31937.1(5)             & M$_{3}$   & 4.81               &  1.66(8) & 1.1\\
32137.4(5)             & N$_{2}$   & 0.28               &  0.59(4) &  -\\
32137.4(5)             & N$_{3}$   & 0.43               &  0.59(4) &  -\\
 \hline
 \multicolumn{5}{l}{$^{a}$The energies of conversion electrons
  are calculated for free atoms using transition} \\
 \multicolumn{5}{l}{
  energies from this table and binding energies from \cite{Sie69}. } \\
 \multicolumn{5}{l}{$^{b}$The intensities of conversion electrons
  are calculated using $\gamma$-ray intensities in this} \\
 \multicolumn{5}{l}{table and conversion coefficients from
  interpolation  in tables \cite{Ros78}. } \\
 \multicolumn{5}{l}{$^{c}$The experimental line widths are taken
                                                from \cite{Pic9a}. } \\
 \multicolumn{5}{l}{$^{d}$The line widths recommended in\cite{Cam01}.
                         $^{e}$Binding energy is taken from \cite{Dra04}. } \\
 \end{tabular}
 \label{tab1}
 \end{center}
\end{table}
 The natural widths of conversion electron lines are also shown in
the table as the precision of energy calibration depends on this
parameter. Particularly, the natural width of the main calibration
line K-32 amounts to 2.83(12)~eV. The conversion lines
N$_{1,2,3}$, due to their sharpness, are well suited for the study
of the spectrometer response function. Unlike the 32.2~keV
transition, which comes only from decay of the isomeric state, the
feeding of the 9.4~keV transition is composed of three branches: a
direct branch from electron capture on the 9.4~keV level, from
electron capture on nuclear levels above the isomeric state and
from the isomeric state via the 32.2~keV transition. Consequently,
even in case of zero retention of $^{83m}$Kr in solid
$^{83}$Rb/$^{83\rm{m}}$Kr source, the 9.4~keV $\gamma$-rays and
corresponding conversion electrons are emitted from the solid
source (the half-life of the 9.4~keV level is 147~ns only which is
deeply below the time needed by the krypton atoms the leave the
source volume). The value of krypton retention in a particular
solid source can be determined from the ratio of the measured 32.2
and 9.4~keV $\gamma$-ray intensities. The 100~\% $^{83\rm{m}}$Kr
retention, i.e. the maximum possible number of isomeric states in
the source, corresponds to the situation when $^{83\rm{m}}$Kr and
$^{83}$Rb nuclei in the source appear in an equilibrium determined
by their decay half-lives. The dependence of the intensity ratio
on the retention deduced from the $^{83}$Rb decay data is shown in
Fig.~\ref{retention}.
 Compiled nuclear data about $^{83}$Rb decay can
be found in \cite{Tul01}.

For $^{83\rm{m}}$Kr atoms in the solid
$^{83}$Rb/$^{83\rm{m}}$Kr source the kinetic energy of conversion
electrons measured by an electron spectrometer is given by the equation
\begin{equation}
  E_{kin}(i) = E_{\gamma} - E_{b}^{solid}(i) - E_{rec}(i) - \phi_{sp},
\end{equation}
where \\
\begin{itemize}
\item{$E_{\gamma}$ is the transition energy which is equal to the energy
of related $\gamma$-rays,}
\item{$E_{b}^{solid}(i)$ represents the binding energy of electron in
subshell $i$ referred to the Fermi level for krypton atoms bound
in the source, }
\item{$E_{rec}(i)$ is the $^{83}$Kr recoil energy encountered after emission of a
conversion electron, }
\item{$\phi_{sp}$ is the spectrometer work function.}
\end{itemize}
Formula (1) is valid for the case when the  source is in electric
contact with the spectrometer. The recoil energy for K-32
electrons emitted from free krypton atoms is of 0.12~eV. The recoil
energy for the other subshell conversion electron is larger. The
largest value is observed for the emission of the N-32 conversion
electrons and amounts to 0.22~eV. Possible changes of these recoil
energies for krypton atoms bounded in the source are very unlikely
at room temperature. The recoil energy after $\gamma$-ray emission
is not stated in this formula as it is negligible. In case of free
krypton atoms this recoil energy amounts to 0.01~eV. Energies of
nuclear $\gamma$-rays $E_{\gamma}$ are known to be stable except
extremely small M\"{o}ssbauer shifts fully negligible in our case.
However, the electron binding energy represents a problem as it
depends on the stability of the physico-chemical environment of
the atom, see Sect.~\ref{23shift}.
The spectrometer work function $\phi_{sp}$ can also
change with a time. Its stability depends both on the spectrometer
surface treatment and on the presence of rest water in
spectrometer vacuum.

\subsection{High resolution electron spectroscopy}
\label{22highres} The interpretation of the conversion electron
spectra measured at high energy resolution requires to account for
two important effects: shake up and off processes connected with
the atom from which the electron originates, and energy losses
resulting from inelastic interaction of the outgoing electron with
surrounding atoms. Both effects produce satellite lines or
continua on the low energy side of the zero-energy loss line,
which is thus reduced in its intensity.

When a conversion electron is ejected from an atom a sudden change
of electric potential of the atom can excite other bound electrons
of the same atom from its orbital either into another (shake up
process - discrete line spectrum) or even into the continuum
(shake off process - continuous spectrum with edge). Actually, in
a shake off process two electrons are  emitted forming two
complementary continuous spectra, one at very low energy and a
second close to the zero-energy loss line. Similar shake up/off
processes were studied experimentally and theoretically also for
$\beta^{+}$ decay, photoelectric effect or electron impact
interaction. In works \cite{Pic9a,Sie69,Car73} and \cite{War91}
shake up/off effects in Kr were observed and interpreted. Shake
up/off effect is not yet fully understood, e.g. in case of the
K-14.4 $^{57}$Co conversion electrons the expected shake up
satellites were not identified. The solid-source effect could mask
the phenomenon in this case, see \cite{Por71}.

The second effect is caused by inelastic scattering of electrons
on surrounding atoms. Namely, even a solid conversion electron
source, prepared from no-carrier added radioactive substance
contains unavoidable traces of contaminants. The radioactive atoms
are deposited on a backing where moreover, electron backscattering
takes place. Finally, a contamination over-layer composed mainly
of carbon, oxygen and hydrogen is formed on the source surface
during and after its production even if it is prepared and kept at
UHV conditions (10$^{-10}$~mbar). Therefore, emitted conversion
electrons suffer from inelastic and elastic scattering processes
in the source material. As a result, components corresponding to
inelastically scattered electrons and zero-energy loss ones appear
in the measured spectra taken with high instrumental resolution.
The energy loss electron spectrum is then represented by a broad
peak, separated from the narrow zero-energy loss electron peak.
The tail of the energy loss distribution extends to zero energy,
see e.g. \cite{Sie65}. The electron scattering process can be
successfully described using Monte Carlo calculations if enough
information about the source is available see, e.g.
\cite{Spa90,Shi92,Fer96}.

\begin{figure}[ht]
 \begin{center}
 \includegraphics[bb=24pt 351pt 378pt 625pt, width=12cm]{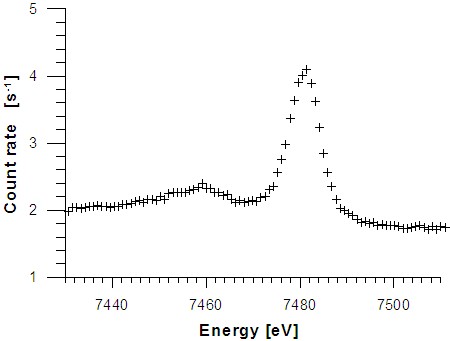}
 \caption[fig3]{Spectrum of L$_1$-9.4 electron conversion line of
 $^{83}$Rb/$^{83\rm{m}}$Kr source No.~15 measured with electrostatic
 spectrometer ESA12 set to instrumental
 resolution of 3~eV.}
 \label{spe015}
 \end{center}
\end{figure}

At a high energy resolution, 3~eV or better, the zero-energy loss
electron peak is well resolved from the shake up/off and loss
energy structures and its energy can be established with high
reliability. The width of this peak is practically represented
only by the convolution of the Lorentzian function (the width of
which is given by the sum of the natural widths of atomic and
nuclear levels in question) with the spectrometer response
function. Fig.~\ref{spe015} shows an example of such a spectrum
which represents a typical result of the measurements at ESA12
spectrometer for one of the $^{83}$Rb/$^{83\rm{m}}$Kr sources
(altogether we investigated 22 sources which are listed in
Table~2).
In contrast, the measurement at lower resolution (tens of~eV)
where the inelastically scattered electrons are not or only partly
resolved from zero-loss electrons, gives the peak energy shifted
to a lower value. Using the material with low Z for the
substrate and carrier free radioactivity reduces the amount of the
inelastically scattered electrons. Conversely, in unfavorable
cases a large contamination in the source material can suppress
the rate of the zero-energy loss electrons substantially, so
that precise energy determination is impossible despite high
instrumental resolution.

\subsection{Chemical and solid state shifts}
\label{23shift} The chemical and/or solid state shifts are defined
as the difference between electron binding energy for free atoms
(obtained in measurement with atoms in a gaseous or vapor state)
and the atom in a particular chemical state and/or solid
environment. The measurements of the shifts represent the main
subject for the specimen investigations by means of the X-ray
photoelectron spectroscopy. The chemical shifts amount to 2--7~eV
with both signs and are correlated with the valence state
\cite{Sie65,Sie69}. For solids, the binding energies are mostly by
about 2--10~eV smaller than for free atoms \cite{Shi77},
\cite{Wei82}. The photoelectron lines are generally found to be
broader for the solid state specimens than for free atoms. A
similar shift of the electron binding energy and consequently a
shift of kinetic energy is observed in spectra of conversion
electrons observed in radioactive decays. In the photoelectron
measurement, the photoelectron reflects the state of specimen
being in a stable chemical and solid state. On the other hand, in
the internal conversion process following $\beta$-decay, in which
the atomic number was changed, the original inner atomic shells
are rearranged during typically 10$^{-16}$~s \cite{Sie65}. As the
half-lives of the nuclear levels are in the order of 10$^{-13}$~s
or longer, the electron binding energy in daughter nucleus atom
must be taken into account, i.e. the chemical and solid state
properties of this atom influence the shift. For reference for the
non zero chemical and solid state shift for the conversion
electrons, see e.g.
 \cite{Por71,Fis85,Kov98}.

For the investigated solid $^{83}$Rb/$^{83\rm{m}}$Kr source, the
$^{83}$Rb activity of 5-10~MBq in a circle of about $\phi$ =
8~mm is needed. It corresponds to an average thickness of only
0.2-0.4 monolayers of radioactive rubidium. After the decay, the
rubidium atoms, originally in the form of oxide or hydroxide in
the source \cite{Leb06}, convert into chemically inert krypton
atoms kept fully or partially in the source solid matrix.
According to the arguments given above, we can expect here only
the effect of the solid state shift.  Thus, the reproducibility
and stability of the conversion electron energy will be determined
by the reproducibility and stability of the source matrix. It has
to be stressed that the values of typical shifts for solids cited
above were observed in the studies with well defined macroscopic
samples. A broader conversion line can be also observed in
comparison with those expected for free $^{83\rm{m}}$Kr atoms.

\section{Instrumentation}
 \label{3instr}
For the development of the solid $^{83}$Rb/$^{83\rm{m}}$Kr source
dedicated instrumentation was necessary: krypton gas target
for $^{83}$Rb production, technique of pure radiochemistry, X-ray and
$\gamma$-spectrometers, vacuum evaporating device and electron
spectrometers for the test of reproducibility and stability of the
electron conversion line energies.

\subsection{Krypton gas target}
 \label{31target}
\begin{figure}[h]
 \begin{center}
 \includegraphics[bb=13 43 796 579 pt clip=true,
  width=12cm, totalheight=8cm]{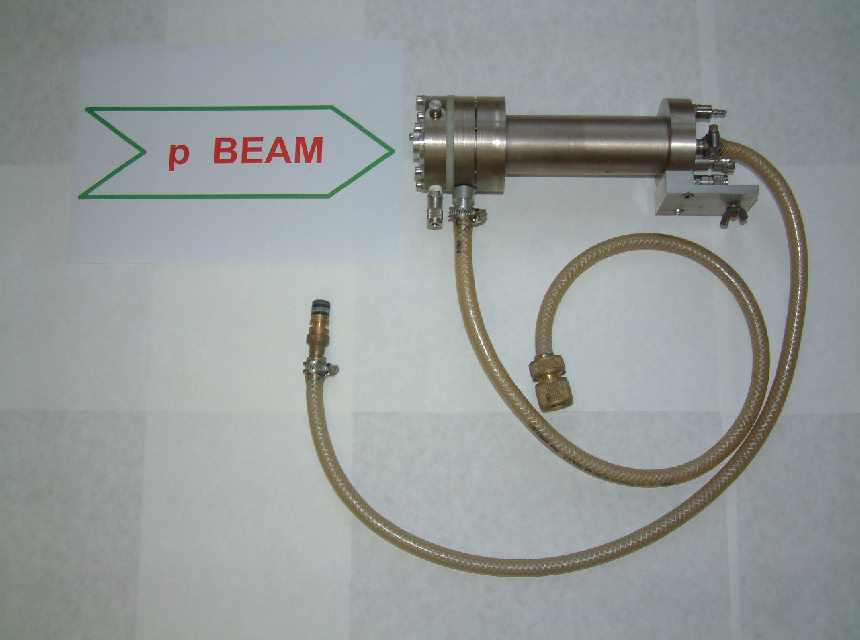}
 \caption[fig4]{The krypton gas target with flexible tubing for cooling
                water.}
 \label{target}
 \end{center}
\end{figure}

\begin{figure}[h]
 \includegraphics[56pt,166pt][441pt,388pt]{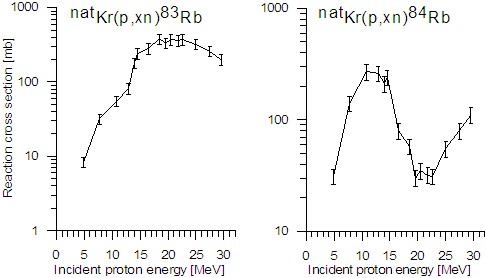}
 \caption[fig5]{Excitation function for the formation of $^{83}$Rb
  and $^{84}$Rb in proton induced reaction on natural krypton taken
  from \cite{Kov91}.}
 \label{yield}
\end{figure}

\begin{figure}[ht]
 \includegraphics[31pt,155pt][410pt,435pt]{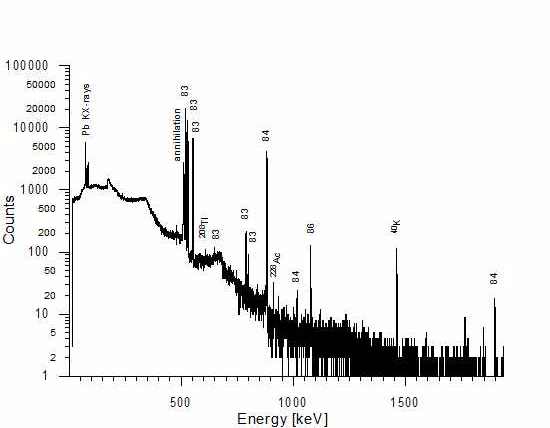}
 \caption[fig6]{$\gamma$-spectrum of the rubidium water solution
 sample measured 600~s at 6\% dead time. The $^{83,\ 84,\ 86}$Rb
 $\gamma$-lines are marked by the numbers 83, 84 and 86,
 respectively. The isotope symbols are used for the lines from the
 background.}
 \label{spek}
\end{figure}

     The production of $^{83}$Rb was carried out at the
U-120M cyclotron of NPI in \v{R}e\v{z} with proton beam via the
reaction $^{\rm{nat}}$Kr(p,xn)$^{83}$Rb using a water cooled krypton
gas target. The basic features of its design and construction
were published in \cite{Kov91}.
Most of target details were produced from a hard aluminium,
free of copper. The entrance window for the beam was from
0.1~mm thick titanium foil. The inner side of krypton vessel was
plated by nickel. For filling with krypton, the target was
equipped with two SWAGELOK miniature quick-connects. In
Fig.~\ref{target}, the overall view on target is shown.
The pressurized krypton gas (absolute pressure of 7.5~bar at room
temperature in volume of 22~cm$^3$) was exposed to the external
6~$\mu$A proton beam for 12 hours (total beam charge of ~250~mC).
During target irradiation, the pressure and temperature increased
to about 9.5~bar and 100~$^\circ$C, respectively. The primary
energy of the proton beam was set to 27.0~MeV. Due to the energy
degradation in cyclotron aluminium beam-focus-smearing foils and
output aluminium window, target titanium entrance window and krypton
gas itself, the energy used for the production was in the
range 19.5--24.1~MeV. This range is optimal for the $^{83}$Rb
production, both maximizing the $^{83}$Rb production rate and
minimizing the amount of $^{84}$Rb (see the excitation functions
in Fig.~\ref{yield}).
The irradiated target was left for a week to let short lived
activities to decay. The mixture of rubidium isotopes deposited
on the target walls was two times washed out
by $\approx$25~cm$^3$ of distilled water. The
produced $^{83}$Rb activity was about 100~MBq and the elution
efficiency of $^{83}$Rb from the target amounted to about 90~\%. The
$\gamma$-spectrum of the rubidium water solution sample measured 14
days after the irradiation is shown in Fig.~\ref{spek}. Only the
$\gamma$-lines from longer living rubidium isotopes and terrestrial
background are present in the spectrum.

As expected, the rubidium water solution contained an admixture
of the rubidium isotopes with mass numbers A = 84 (T$_{1/2}$ =
32.9~d) and A = 86 (T$_{1/2}$ = 18.66~d). The relative abundance
of isotopes $^{83}$Rb, $^{84}$Rb and $^{86}$Rb amounted to 10:1:0.2
at 14th day after the irradiation.
The abundance ratio for $^{83}$Rb and $^{84}$Rb was in agreement
with the measurement \cite{Kov91}, see Fig.~\ref{yield}. These
admixtures are relatively weak and did not cause any problem for
the measurement of the low energy conversion electrons. The amount
of $^{86}$Rb was in fact negligible due to its low yield
 and relatively short half-life. Even higher yield and better
radionuclide purity could be reached using enriched $^{84}$Kr in the
target instead of natural krypton since the contribution due to
the reactions $^{86}$Kr(p,n)$^{86}$Rb, $^{86}$Kr(p,3n)$^{84}$Rb
and $^{83}$Kr(p,$\gamma$)$^{84}$Rb would be avoided. Some other
aspects of $^{83}$Rb production related to the KATRIN project can
be found in \cite{Ven05}.

\subsection{Pure radiochemistry and production of the
$^{83}$Rb/$^{83\rm{m}}$Kr solid source at vacuum evaporation
device}
 \label{32evap}
The $^{83}$Rb/$^{83\rm{m}}$Kr sources of conversion electrons were prepared
by means of vacuum evaporation. In order to minimize  electron
energy loss, the $^{83}$Rb had to be deposited onto a metal
evaporation boat in the cleanest possible form. For this the water
rubidium solution obtained after gas target wash out had to be
reduced in volume and purified. First, the solution was
gradually evaporated under an infra red-lamp in a quartz vessel. Then
the activity from the quartz vessel was washed out several times
into the total amount of $0.5\div1.5$~ml of 0.1~M HNO$_{3}$ acid.
This primary eluate was let to pass through a chromatography
micro-column (70~mm long and 1.5~mm in diameter). To speed up the
column flow nitrogen gas with a purity of 99.99~\% and pressure of 1.7
bar was applied to the column. Elementary rubidium was retained in
the upper part of the column. Next, 0.3, 0.5 and 1.0 molar
solutions of HNO$_{3}$, each of 0.5~cm$^{3}$ volume, were applied
to purify the column. Rubidium was released from the column by
means of a 0.5~cm$^{3}$ of 2.0~M HNO$_{3}$. The acid was
removed from this last eluate by evaporation. By dissolving the
rest in distilled water, the solution for the boat deposit was
obtained. Successively, three ion-exchange catex resins were
tested for filling of the column --- OSTION LG KS 0803 (with
particle size of 15--20 $\mu$m), AMINEX A4 (16--24 $\mu$m) and
LEWATIT LWC-100 (1--60 $\mu$m) with practically the same result.

In addition a more simple and faster process of purification was
suggested and tested. Instead of chromatography column, the
primary eluate was purified using 220 nm pore aerosol filter from
MILLIPORE and then evaporated to a small volume for boat
deposition. Unfortunately, solid $^{83}$Rb/$^{83\rm{m}}$Kr sources
prepared in this way, exhibited imperfections and this procedure
had to be excluded from further source development, see
Sec.~\ref{41rez}.

\begin{figure}[h]
 \begin{center}
 \includegraphics[bb= 14 574 297 786 pt, clip=true,
  width=11cm,totalheight=8cm]{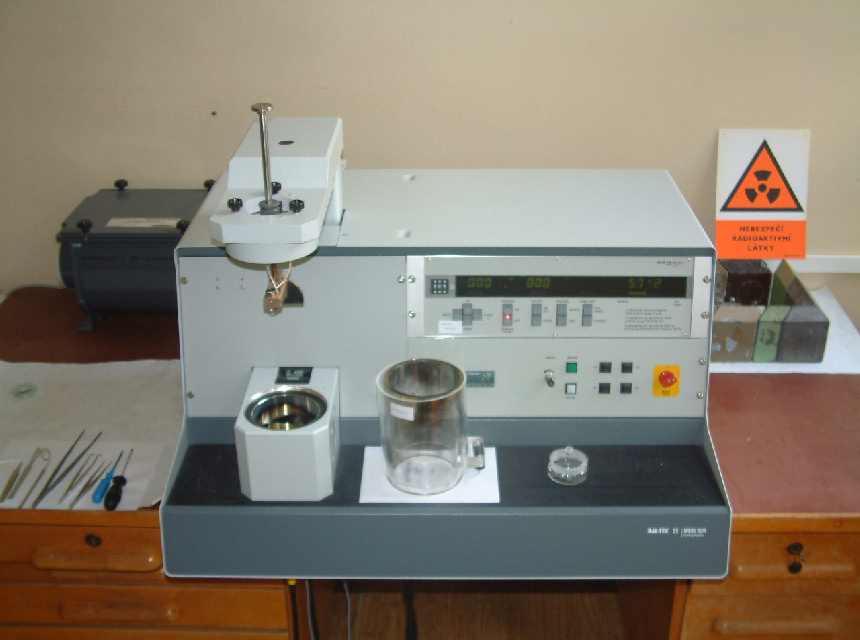}
 \caption[fig7]{The high vacuum evaporation device MED 020.}
 \label{MED020}
 \end{center}
\end{figure}

The modular high vacuum
coating system MED 020 from BAL-TEC (Liechtenstein) was
used for the vacuum evaporation, see Fig.~\ref{MED020}.
The glass cylindrical vacuum chamber with 108~mm inner diameter
and 172~mm height was evacuated with an oil free pumping system
consisting of a fore-vacuum membrane pump (2~m$^{3}$/h) and a
turbomolecular pump (70~l/s). The system enabled us to reach
a vacuum of 3 $\times$ 10$^{-6}$~mbar in evaporation chamber.
To reduce the contamination of the chamber by $^{83}$Rb a
removable protection foil of 0.2~mm thick aluminium was used to line
on the inner wall of the cylindrical part of the chamber before
each evaporation process. A Microfibre HEPA filter was connected to
the vacuum system outlet to prevent any escape of evaporated
radioactivity. The radioactive material was electrically heated in
a metal boat using a high current supply (4~V; maximal 250~A). In
most cases, tantalum boats were used, but molybdenum and tungsten
ones were tested, too. The dependence of the boat temperature on
the heating electrical current was determined by finding the
current value at which a particular  metal melts. Indium
(157~$^\circ$C), tin (232~$^\circ$C), aluminium (660~$^\circ$C),
silver (962~$^\circ$C) and gold (1064~$^\circ$C) were used. The
calibration depended on the material and dimension of a given
evaporation boat; e.g. gold melted at $\approx$30~A (using 4~V~\
current supply) in a tantalum boat (37~mm length, 12~mm width,
0.095~mm thickness)  used in the majority of the last $^{83}$Rb
evaporations. To check the dependence of  $^{83}$Rb evaporation on
the temperature, a number of boats were heated to gradually
increased temperatures. By means of $\gamma$-spectroscopy the
activity remaining in the boats was checked after the heating.
A decrease of activity was not found up to 300~$^\circ$C. In these
cases where the boats were heated above 600~$^\circ$C, only a few percent
of  $^{83}$Rb activity remained in the boat. The studied
$^{83}$Rb/$^{83\rm{m}}$Kr vacuum evaporated sources were prepared
according to \cite{Kov93}. To avoid some possible impurities the
boat with $^{83}$Rb activity was preliminary heated to
200~$^\circ$C for 5--10 minutes. In this phase the boat was
shielded with special shutter to keep the substrate clear. Then
the $^{83}$Rb evaporation followed for 30--60 seconds at 800~$^\circ$C.

\begin{figure}[h]
 \begin{center}
 \includegraphics[bb= 11 573 297 786 pt, clip=true,
  width=10cm,totalheight=7cm]{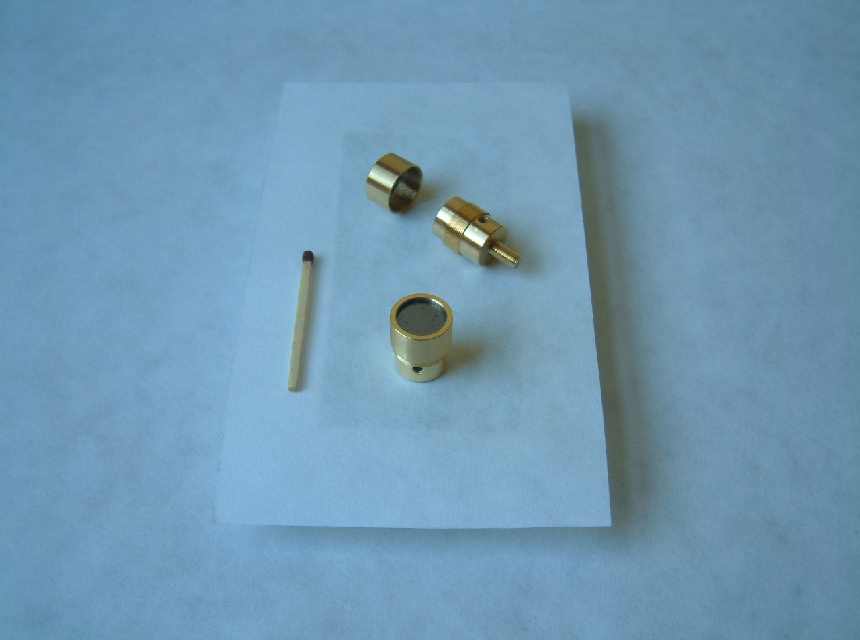}
 \caption[fig8]{The $^{83}$Rb/$^{83\rm{m}}$Kr source prepared by
                vacuum evaporation on 0.2~mm thick carbon foil
                substrate and fixed
                in a copper gold plated holder.}
 \label{source}
 \end{center}
\end{figure}

The vacuum evaporated source was of circular shape with 8~mm
diameter. It was determined by a circular mask with thickness
of 1~mm situated directly
on the source backing during the evaporation process. A 50~$\mu$m
thick aluminium foil or carbon (0.2~mm foil or 1~mm thick disk of
highly oriented pyrolytic graphite (HOPG)) served as a $^{83}$Rb
source backing. The distance between the source backing and the
evaporation boat was from 17 to 1~mm. During heating to
800~$^\circ$C the tantalum boat was deformed due to thermal
extension. The deformation in the direction normal to the
boat plane could reach 2~mm causing mostly a reduction of the distance
between the boat dimple (2~mm deep) with $^{83}$Rb and the source
backing. For larger boat-source backing distances
($>$~10~mm) the evaporated source activity was nearly
2$\times$ higher than the $\Omega/2\pi$ of initial activity in the
boat (solid angle $\Omega$ was determined by the boat center point
and an 8~mm diameter circular mask). At smaller distances the
evaporation efficiency increased but it was less than
$\Omega/2\pi$. For further manipulation and for the measurement at
spectrometers the source was fixed in a copper holder plated with gold
shown in Fig.~\ref{source}. The gold on the holder ensured safe
electric contact of the source backing to the spectrometer zero potential.

\begin{figure}[h]
 \begin{center}
 \includegraphics[bb= 13 14 222 234 pt, clip=true,
  width=8cm,totalheight=8cm]{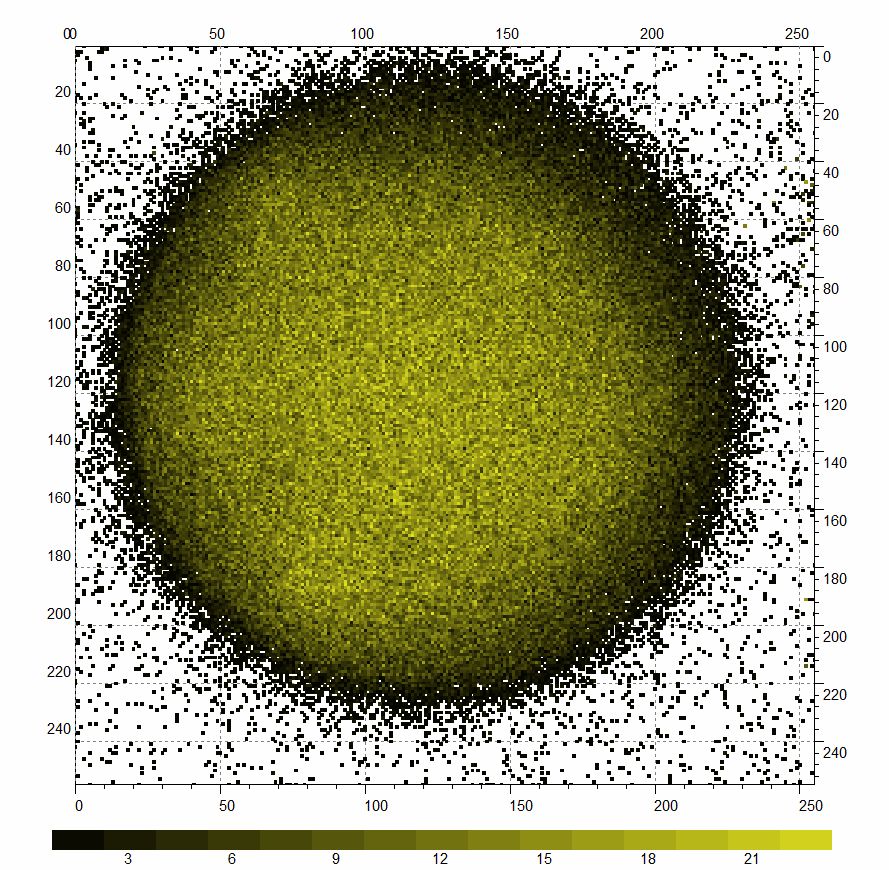}
 \caption[fig9]{The Timepix image of $^{83}$Rb/$^{83\rm{m}}$Kr source
                No.~9. The full scales correspond to 14.08~mm.}
 \label{timepix}
 \end{center}
\end{figure}

Distribution of $^{83}$Rb/$^{83\rm{m}}$Kr activity on two of
vacuum evaporated sources was examined with a position and energy
sensitive detector of the Timepix type \cite{Llo07}. The
measurements were carried out in the Institute of Experimental and
Applied Physics of the Czech Technical University in Prague. The
detector has 256 x 256 pixels and a sensitive area of 14.08~mm x
14.08~mm. Dimensions of individual pixels are 55~$\mu$m x
55~$\mu$m implying outstanding position resolution. The thickness
of the Timepix silicon chip amounts to 300~$\mu$m. In order to
prevent damage or contamination of the detector surface a 4~$\mu$m
mylar foil was placed between the chip and the source fixed in a
holder. Since the $^{83}$Rb and
$^{83\rm{m}}$Kr decays are accompanied by X-rays, gamma rays,
internal conversion- and Auger electrons, the interpretation of
the measurements is not straightforward. In particular, intensive
gamma-rays with energy around 500~keV can produce a signal in
several pixels especially if their direction is not perpendicular
to the detector chip plane. The Timepix image of the
$^{83}$Rb/$^{83\rm{m}}$Kr source No.~9 projected on the horizontal
plane is shown in Fig.~\ref{timepix}. The distance between the
source and chip plane was 1~mm. The activity of the source was
26~kBq at the time of measurement. No distinct inhomogeneities
exceeding 55 micrometers were recorded. Details can be found in
\cite{Ven07}.

\subsection{The $\gamma$-detectors}
 \label{33gamma}
\begin{figure}[h]
 \begin{center}
 \includegraphics[bb= 14 574 297 786 pt, clip=true,
  width=11cm,totalheight=8cm]{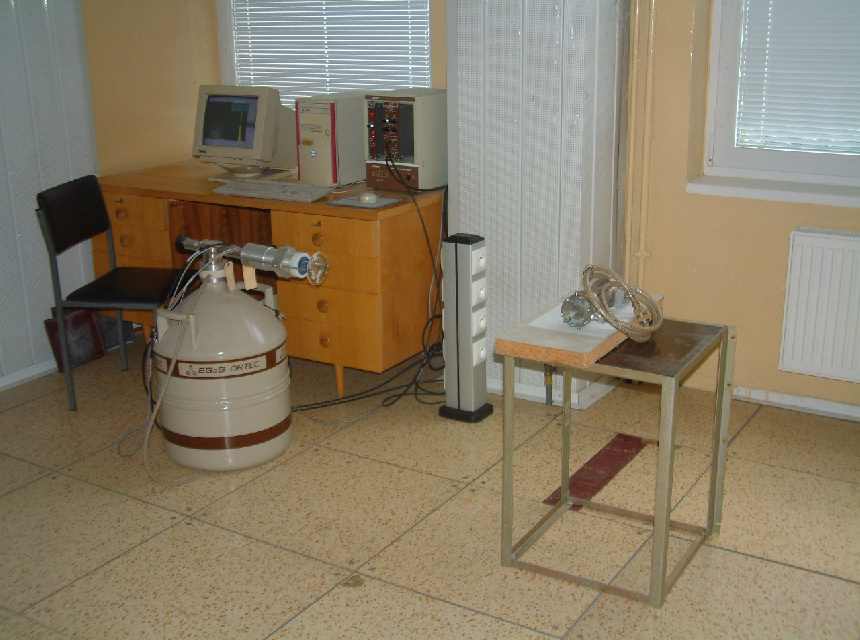}
 \caption[fig10]{The measurement of the irradiated gas target using
                HPGe $\gamma$-spectrometer.}
 \label{HPGe}
 \end{center}
\end{figure}

\begin{figure}[h]
 \begin{center}
 \includegraphics[clip=true, bb=13 573 297 786pt, width=11cm]{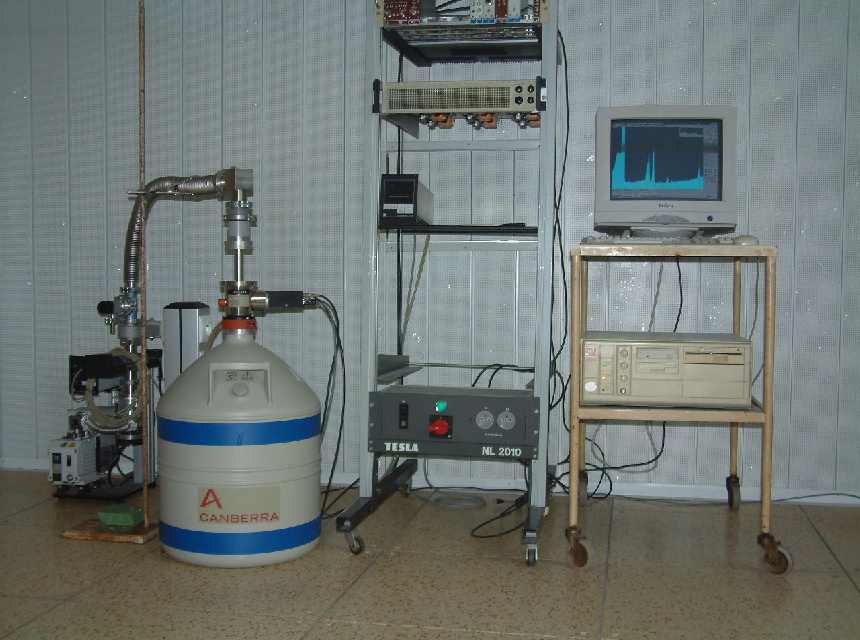}
 \caption[fig11]{The measurement of the $^{83}$Rb/$^{83\rm{m}}$Kr sample in
                vacuum chamber by means of the Si(Li) X-ray spectrometer.}
 \label{SiLi}
 \end{center}
 \end{figure}

\begin{figure}[h]
 \begin{center}
 \includegraphics*[bb= 20 355 406 652 pt, clip=true,
  width=12cm,totalheight=8cm]{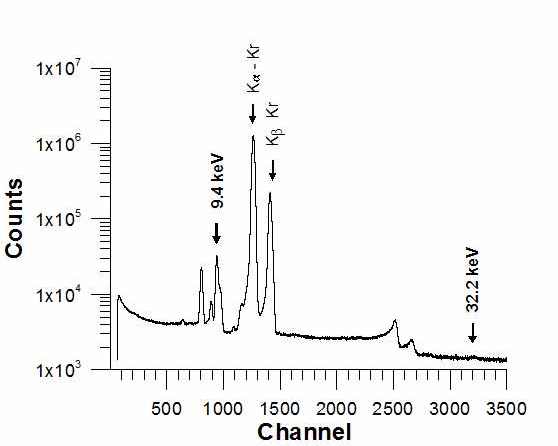}
 \caption[fig12]{The $\gamma$-spectrum of $^{83}$Rb/$^{83\rm{m}}$Kr
 source No.~28 measured with the Si(Li) detector. The $\gamma$ and X-rays
-lines from the decay are marked. The other lines represents Cu and Au X-rays
induced in the source holder, detector silicon escape X-rays and summing peaks.}
 \label{SiLispectrum}
 \end{center}
\end{figure}

\begin{figure}[h]
 \begin{center}
 \includegraphics*[bb= 18 348 386 599 pt, clip=true,
  width=12cm,totalheight=8cm]{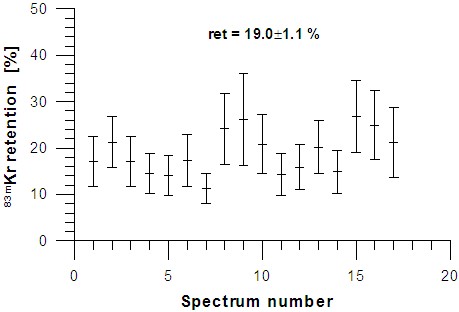}
 \caption[fig13]{The result of 5 days retention measurement for
the source No.~28. The individual values of retention are
displayed in dependence on the spectrum number. Resulting mean
value of retention $ret=19.0(11)$~\% is indicated.}
 \label{retent28}
 \end{center}
\end{figure}

 The radiopurity and activities of the radioactive samples
and the retention of $^{83\rm{m}}$Kr in $^{83}$Rb/$^{83\rm{m}}$Kr
solid sources were determined via $\gamma$-spectrometry based on
a commercial apparatuses. The $\gamma$-ray spectra were measured with
a 20\% efficiency HPGe detector equipped with standard preamplifier, amplifier
and computer based ADC card TRUMP --- all from EG\&G
ORTEC. The energy resolution of the system, the full width at half
maximum, was 1.8~keV at 1332~keV energy. The spectroscopy chain was
calibrated for energy and absolute efficiency by a set of
commercially available standard radioactive sources of $^{133}$Ba,
$^{152}$Eu and $^{241}$Am. The activities of rubidium isotopes were
calculated from the net peak areas corrected for
the dead time and efficiencies with the use of the relevant
$\gamma$-intensities published in \cite{Fir98}. In view of the large
span of sample activities, the spectrometer absolute efficiency
calibrations for distances source-detector endcup 50, 160 and 320~cm
were established
in order to achieve optimal dead time for the measurements. In
Fig.~\ref{HPGe}, the view on the $\gamma$-spectrometer with the krypton
gas target is shown.
The $^{83\rm{m}}$Kr retention in $^{83}$Rb/$^{83\rm{m}}$Kr solid
samples was determined from the measurement of the $\gamma$-ray
intensity ratio for the 32.2 and 9.4~keV transitions according to
Sec.~\ref{21conv}. For this purpose, a Si(Li) detector equipped
with beryllium window was employed as the HPGe $\gamma$-detector
was practically insensitive for the detection of these low energy
$\gamma$-radiations. A 80~mm$^{2} \times $5~mm Si(Li) detector
with build-in preamplifier and spectroscopy amplifier (both
devices from CANBERRA) were
used. The thickness of the beryllium window was 0.05~mm. The
card TRUMP digitized the amplifier output signals into 8192
channels. The energy resolution of the system was 330~eV at an energy
of 33~keV. Energy and absolute efficiency calibrations of the
detector were based on calibrated sources of $^{55}$Fe and
$^{241}$Am. The calibration was carried out with the KX, LX and
$\gamma$-ray energies and intensities from \cite{Fir98} and
\cite{Lep94}.
 The X-ray spectrometer was equipped with a small vacuum
chamber (content of about 100~cm$^{3}$) pumped by vacuum station
from LEYBOLD. The station was assembled from a two stage
rotary pump (3 m$^{3}$/hour) and turbomolecular pump (50~l/s). The
end vacuum achieved was of 3 $\times$ 10$^{-6}$~mbar. The 1.5~mm thin
plexiglass window on the vacuum chamber facing to the detector
with attenuation factors of 0.49 and 0.98 for the 9.4 and 32.2~keV
$\gamma$-rays, respectively, allowed us to measure the
$^{83\rm{m}}$Kr retention for the $^{83}$Rb/$^{83\rm{m}}$Kr
samples in vacuum. In Fig.~\ref{SiLi}, the view on the
$\gamma$-spectrometry measurement with the Si(Li) detector and vacuum
chamber is shown.
Fig.~\ref{SiLispectrum} shows as an example the gamma spectrum of the
$^{83}$Rb/$^{83\rm{m}}$Kr source No.~28 measured with Si(Li)
detector for the retention determination. At a distance between
source and detector endcup of 23~mm the dead time amounted to 7~\%.
The time of the measurement was set to 6~h.
The analysis of retention measurement is demonstrated
in Fig.~\ref{retent28}. For 17 spectra subsequently measured with
source No.~28 the values of retentions were established.

\subsection{Electrostatic spectrometer ESA12 at NPI \v{R}e\v{z} }
 \label{34esa12}
The energy stability of electrons from solid rubidium sources was tested
with the ESA12 spectrometer at NPI \v{R}e\v{z} with the
L$_{1}$-9.4 conversion line at an energy of 7.5~keV. The monitoring K-32
line at energy 17.8~keV could not be tested as it falls out of
the high resolution range ($\le 3~$eV) of this instrument.
The result of such tests will help to plan more effectively the
K-32 measurements at a less accessible electron
spectrometer at Mainz university. The long-term stability tests
lasting many weeks were more easily carried out at
home \v{R}e\v{z} workplace.
They are also much cheaper since no liquid helium for
superconducting magnets is needed.

\begin{figure}[h]
 \begin{center}
 \includegraphics*[bb= 14 502 225 786 pt, clip=true,
  width=8cm,totalheight=11cm]{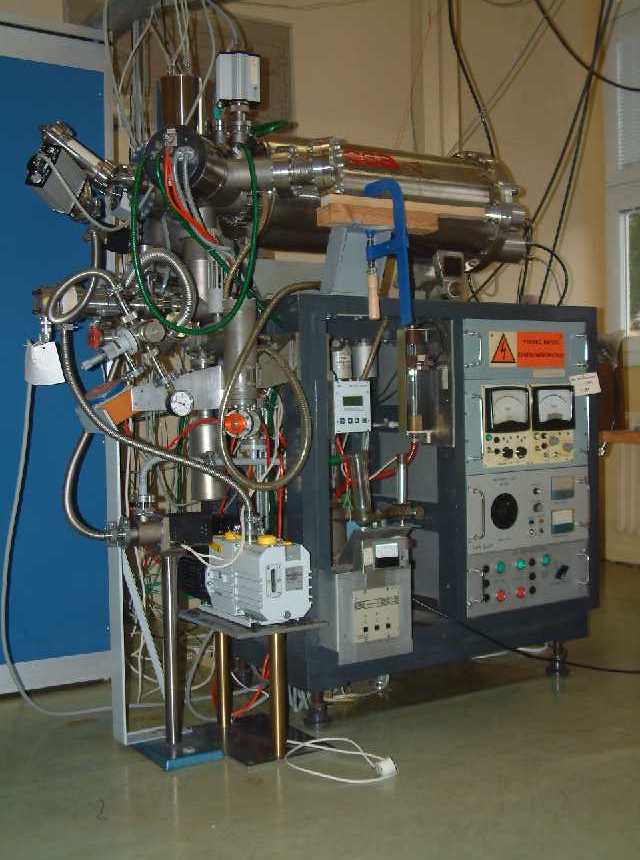}
 \caption[fig14]{Electron electrostatic spectrometer ESA12 at NPI \v{R}e\v{z}.}
 \label{ESA}
 \end{center}
\end{figure}

The ESA12 spectrometer in \v{R}e\v{z} is a second order focusing
cylindrical mirror electron analyzer equipped with
retarding/accelerating lens system and channel electron multiplier
as a detector \cite{Var82,Dra95}. In Fig.~\ref{ESA}, the view on
the spectrometer with vacuum components and 220 V power
distribution is shown.
The primary value of the kinetic energy for electrons detected
at ESA12 $E_{k}$ is given by the relation
\begin{equation}
  E_{k} = E_{lens}+E_{an}=eU_{lens}+(1/c_{an})\times|U_{an}|,
\end{equation}
where
\begin{itemize}
\item{ $E_{lens}$ is the energy lost or gained at lens,}
\item{$E_{an}$ is the analyzer passing energy,}
\item{$e$ represents a unit of elementary charge,}
\item{ $U_{lens}$ is
electric potential on the lens electrode connected electrically with
the radioactive source,}
\item {$c_{an}=0.56$~V/eV is a constant determined by
the analyzer construction,}
\item{ $U_{an}$ is negative voltage
applied to the outer analyzer cylinder.}
\end{itemize}
The behavior of $E_{k}$, i.e. its stability and reproducibility,
is the most important parameter followed in the
$^{83}$Rb/$^{83\rm{m}}$Kr solid source development. The
instrumental resolution of the ESA12 is given by the relation
$\Delta E = 0.011 \times E_{an}$ and its response function is
assumed to have a Gaussian shape. The spectrometer is a
one-channel device which can work in basic or lens mode. In the
first case, the $U_{lens}=0$~V and $U_{an}$ is swept with
adjustable voltage step in order to cover the required kinetic
energy range according to equation (2) (the steps can be
chosen in voltage ascending or descending way). For the lens mode
the $U_{an}$ is stabilized at an appropriate value to provide
the intended analyzer resolution and the $U_{lens}$ is changed
in the same way as $U_{an}$ is changed in the basic mode. Generally accepted method for the
precise measurement of high voltage based on HV resistor divider
is used. Three precision HV dividers named KD2, KD3 and KD10 were
developed in our laboratory \cite{Dra05}. First two are
similar with dividing ratio of $\approx$402.7, the third one has
the dividing ratio of $\approx$9.9.
  The measurement of the actual values of
voltages at the outputs of the dividers is accomplished with
high-quality commercial digital voltmeters. The dividing ratio $d$
of the dividers was chosen in such a way that the voltmeters can
operate in the 20~V range, thus providing the most precise values
up to 8 valid digits. When a voltage step is set, i.e. the energy
point is adjusted, the detector signals are added up in the
counter within a fixed period (formerly 1 now 2 sec). During the
measurement, the
spectrometer input and output slits (situated at the inner
cylinder) are set to the maximal width of 4~mm. In this case and
for the basic mode the geometrical transmission of the system
amounts to 0.74~\% of $4\pi$. In the lens mode, the transmission
decreases or increases with the applied retarding or accelerating
lens voltage. Using lens mode, an excellent resolution can be
obtained also for higher energy electrons by setting a low value
for the passing energy. However, as stated above, the transmission
of the analyzer is reduced due to the loss of electrons when
passing the electrostatic retarding field between lens
electrode and the input analyzer slit (the reduction factor is
roughly proportional to $1/U_{lens}$ for electrons with given
kinetic energy). The total efficiency of the spectrometer is
obtained by multiplying of the transmission with a channeltron
efficiency. The channeltron efficiency is energy dependent and
amounts to several tens of percent for electrons in a range of
10--1000~eV and several percent for the electrons in~keV
range. The low overall efficiency of the spectrometer causes the
measurement to be time consuming --- two days or more with a
$^{83}$Rb/$^{83\rm{m}}$Kr
source of several MBq.

A computer program controls the measurement, i.e. digital
voltmeters attached to HV dividers, precise HV power supplies and
a counter, according to given input data. After all energy
points in a given energy range are scanned the basic element of a
measurement, the sweep, is completed. The sweeps are then repeated
until sufficient statistics is obtained. The number of counts and
the related measured voltage $U_{an}/d$ in basic mode or
$U_{lens}/d$ in lens mode were recorded successively into the
output file for each voltage step and each sweep. In the lens
mode, the actual values of stabilized voltage $U_{an}/d$ were not
recorded in the output file formerly. If the voltage $U_{an}/d$ appeared
outside fixed limits the control program wrote appropriate
comment into the output file.
At present this voltage is recorded as well.

During the measurement, the
control program allows to visualize the electron spectrum sorted from
all already accumulated sweeps on the computer monitor. After finishing the
measurement, the spectrum is sorted off-line using the data from the output file.
In the measurement of L$_{1}$-9.4 conversion electrons with kinetic
energy of 7482~eV the following three basic control program modes
with values for $E_{an}$, energy ranges and steps
were used:
\begin{itemize}
\item{basic mode: \hspace{2.2cm} $E_{begin}=7150$~eV $E_{end}=7665$~eV $E_{step}=5$~eV}
\item{lens mode $E_{an}=260$~eV: $E_{begin}=7350$~eV $E_{end}=7550$~eV $E_{step}=1$~eV}
\item{lens mode $E_{an}=260$~eV: $E_{begin}=7467$~eV $E_{end}=7496$~eV $E_{step}=0.5$~eV}
\end{itemize}
The first two modes include a larger portion of the spectrum on the low
energy side of the L$_{1}$-9.4 line, which is useful for
inspection of the loss electron spectrum. The
third mode is most frequently applied as it is used for the line
position determination. From this it follows that the precise and
reproducible measurement of voltages $U_{an}=-260 \times
0.56=-145.6$~V and $U_{lens}/d \approx 17.9$~V is of decisive importance.

\begin{figure}[h]
 \begin{center}
 \includegraphics*[bb= 13 351 396 596 pt, clip=true,
  width=12cm,totalheight=8cm]{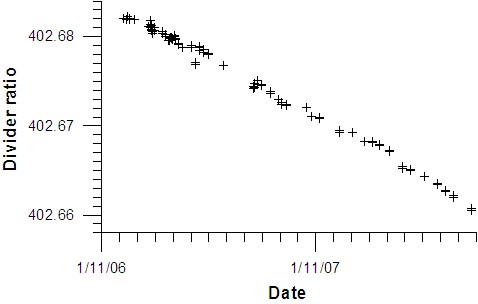}
 \caption[fig16]{Dividing ratio for KD3 measured with +650~V at input
  in the period Dec, 2006--Oct, 2008.}
 \label{ratio3}
 \end{center}
\end{figure}

During the development of the solid $^{83}$Rb/$^{83\rm{m}}$Kr source
a number of improvements was gradually adopted at the ESA12 spectrometer:
\\
a) new voltmeters and one of the HV supplies were installed, \\
b) frequent measurement of the divider ratio was introduced, \\
c) vacuum system was improved, \\
d) open water cooling circuit was replaced by a closed water system, \\
e) auxiliary voltage of --29~V on the channeltron input funnel was applied, \\
f) control program was modified, \\
g) more precise analysis of the measured data was introduced. \\
\\
a) \\
In the old HV system, the voltmeter SOLARTRON 7061 (produced in
1989, set to resolution of 6.5 digits) attached to the divider KD2
or KD3 was used at basic mode for the cylinder electrode and at lens
mode for the lens electrode. The voltmeter TESLA M1T330 (produced
in 1985, 5.5 digits) attached to divider KD10 was used at lens mode for
the analyzer electrode. Both these voltmeters were replaced by two new
ones FLUKE 8508A, both set to resolution of 7.5 digits. The
8508A FLUKE multi-meter is designed for the most demanding
measurement applications and provides extremely high measurement
precision. It is one of the best available on the world market,
intended also for the KATRIN experiment. The measurements with
FLUKE and SOLARTRON devices connected simultaneously to
batteries $3\times4.5$~V or a Weston cell 1.0~V showed that the
ripple and noise of the voltage indications was by factor of 2.7
smaller with the FLUKE voltmeter. The temperature coefficient of
Fluke voltmeters was also measured. For the voltages 1.6, 14.5 and
650~V at ranges 2, 20 and 1000~V values of only
+0.08, +0.1 and --0.07~ppm/$^\circ$C we obtained. The measured
temperature coefficient for SOLARTRON at the voltage 14.5~V
amounted to +0.4~ppm/$^\circ$C. Generally, the voltmeter
indications exhibit a time drift, being larger when the instrument
is new. The FLUKE voltmeter specification shows a 365 days uncertainty
of about 3~ppm at a confidence level of 95\% for DC voltage
measurement. In order to check the drift behavior, one of the
FLUKE voltmeter was sent till now 4 times to The Czech Metrological Institute
(CMI) Brno, for verification of the voltmeter performance at
metrological standard voltages for all DC ranges and at the
voltages important for our measurement, i.e. 1.6, 18, --150 and
650~V (verification does not mean a calibration at which the
voltmeter is reset in order to show the right voltages). The
voltages 18 and 150~V are close to those used in measurement of
the L$_{1}-9.4$ spectrum; the values 1.6 and 650 are related to
the measurement of the dividing ratio (see below). In
Fig.~\ref{kalfluke} the difference between the
voltage measured by voltmeter and those set by precise metrological
divider for the important values is shown on dependence of time.
\begin{figure}[h]
 \begin{center}
 \includegraphics*[bb= 13 343 398 582 pt, clip=true,
  width=12cm,totalheight=6cm]{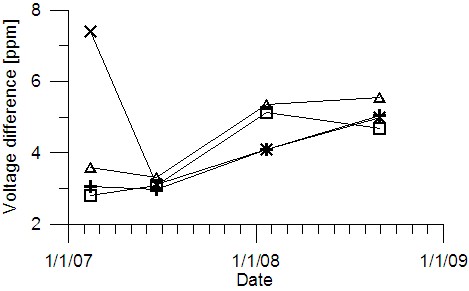}
 \caption[fig15]{Result of 4 successive FLUKE voltmeter verifications at
  CMI --- differences of $U_{meas.}-U_{set}$ for 4 voltages in ppm units:
  ~+1.6, $\Box$~+18, $\bigtriangleup$~ --150 and $\times$~+650~V.}
 \label{kalfluke}
 \end{center}
\end{figure}
The largest drift amounts only to 0.22~ppm/week (between the first and
second verification of 650~V). In the approximation, that the voltmeter
indication changes linearly with time, the
measured energy of the conversion electron line can be corrected for
the voltmeter drift. \\
The HV supply --3~kV ATOMKI (produced in 1989) utilized in
combination with the KD10 divider for $U_{an}$ at the lens mode
measurement was replaced by a MCP~34--650 MOD power supply (maximal
voltage at the output is 650~V, + or -- polarity can be set by
grounding of the appropriate output terminal)
from F.u.G ELECTRONIC. The relative value off p-p
ripple and noise measured at a voltage set to -150~V for the old
power supply amounts to $2.9\times10^{-4}$, whereas for the new
device much a better value of $3.6\times10^{-6}$ is obtained. In
order to reduce the number of electronic elements the divider KD10 was
removed from the system as the FLUKE voltmeter enables us to
measure the analyzer voltage ($0\div-650$~V) with sufficient precision
directly. The second HV supply produced at NPI and used for the
cylinder or lens electrode in basic or retarding/accelerating
modes, respectively, remains unchanged.

b) \\
Besides the precise voltmeter values, a detailed knowledge about
the behaviour of the HV dividers KD2 and KD3 was necessary. Each of them
consist of 74 TESLA TR164 resistors connected in series and assembled
into two groups forming high and low voltage sections. Moreover,
matched pairs of resistors with positive and negative temperature
coefficients were combined in the dividers in order to
reduce the dividers temperature coefficient.
The resistance of high and low sections are 54.013 M$\Omega$ and
134.4 k$\Omega$, respectively.
Further details about dividers can be found in \cite{Dra05}.
Starting December 2006, the
measurements of the dividing ratio with +650~V at the KD3 divider inputs
are frequently carried out, see Fig.~\ref{ratio3}.
By fitting of the line to the dependence in the figure a value
of divider drift of --0.6(1)~ppm/week was determined. The drift measured for the
second divider KD2 in the same way amounts to -1.0(1)~ppm/week.
The temperature coefficient
of both dividers deduced from divider ratio measurements taken at
different temperatures amounts to about --0.2~ppm/$^\circ$C. Again,
if necessary, the measured energy of conversion electrons can
be corrected for the divider drift. Actually, during the
measurement of the electron line L$_{1}$-9.4, a voltage of about 7200~V is
applied to the divider KD3 or KD2. The dividers themselves exhibit at
different input voltage a slightly different ratio $d$.
Nevertheless, we can use the divider ratio determined at 650~V for two
reasons: \\
--- there are no indications that the divider ratio drift itself is different
for the different input voltages \cite{Mar07}. \\
--- for the energy stability of the conversion line
the knowledge of the absolute energy is not necessary. \\
Unfortunately, there is no possibility to measure precisely the
divider ratio at such a high voltage in the Czech Republic.
In order to learn more this question, KD2 divider
was moved to The Institute of Physics
at Mainz University, where the measurement of divider ratio
and its drift at 7200~V was started.

\begin{figure}[h]
 \begin{center}
 \includegraphics*[bb= 13 13 581 830 pt, clip=true,
  width=12cm,totalheight=6cm]{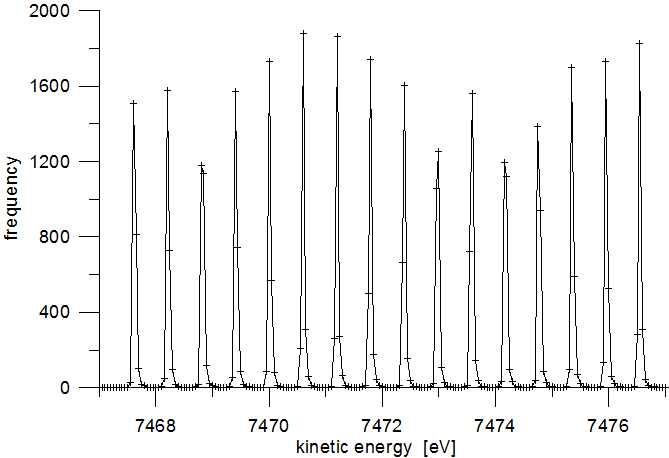}
 \caption[fig17]{D1: frequency distribution of the energy values using bins
                 with width of 0.05~eV. The dependency was obtained for
                 the measurement with the
                 $^{83}$Rb/$^{83\rm{m}}$Kr solid source, line L$_{1}$-9.4.
                 Only the low energy part of dependency is shown .}
 \label{D1}
 \end{center}
\end{figure}

\begin{figure}[h]
 \begin{center}
 \includegraphics*[bb= 13 13 581 830 pt, clip=true,
  width=12cm,totalheight=6cm]{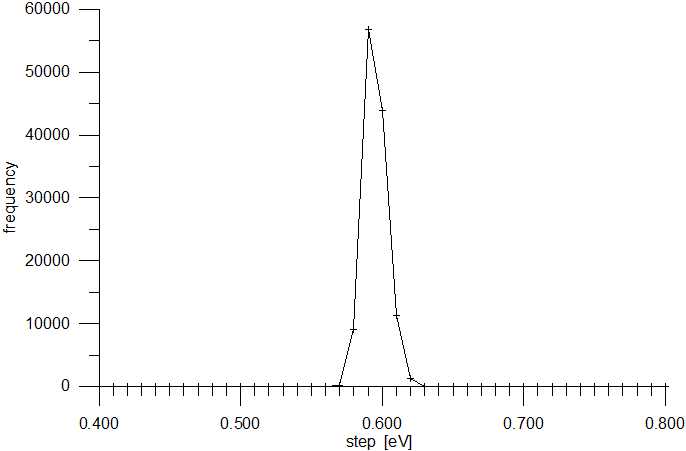}
 \caption[fig18]{D2: frequency distribution of the step values using bin
                 with width of 0.01~eV. The dependency was obtained for the
                 measurement with the $^{83}$Rb/$^{83\rm{m}}$Kr solid source,
                 line L$_{1}$-9.4, where $E_{step}=0.5$~eV was used.
                 Only the range (0.4, 0.8)~eV is shown.}
 \label{D2}
 \end{center}
\end{figure}

\begin{figure}[h]
 \begin{center}
 \includegraphics*[bb= 13 13 581 830 pt, clip=true,
  width=12cm,totalheight=6cm]{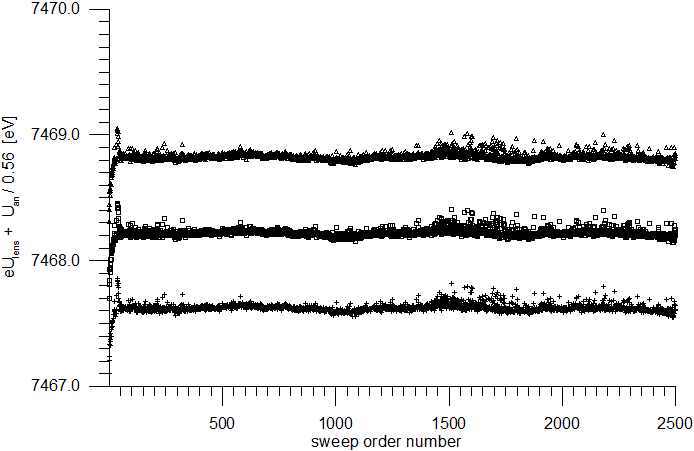}
 \caption[fig19]{D3: dependence of
                 ($eU_{lens}+(1/c_{an})\times|U_{an}|$) value in
                  1-st, 2-nd and 3-rd point on
                 the sweep number.}
 \label{D3}
 \end{center}
\end{figure}

\begin{figure}[h]
 \begin{center}
 \includegraphics*[bb= 13 13 581 830 pt, clip=true,
  width=12cm,totalheight=6cm]{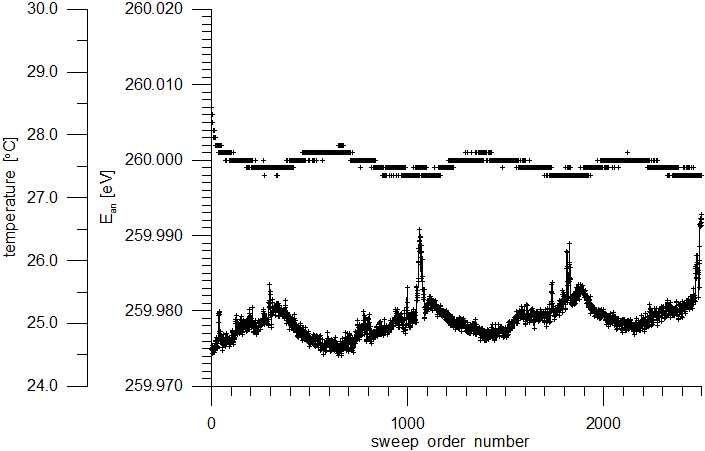}
 \caption[fig20]{D4: dependence of ($(1/c_{an})\times |U_{an}|$) value
                 (upper graph) and temperature (lower graph), averaged
                  for all points in sweep, on the sweep number.}
 \label{D4}
 \end{center}
\end{figure}

A 650~V at the divider input represents a rather low voltage at
the divider output, of about 1.6~V. In this case, the divider
output connectors, voltmeter terminals and connecting leads
represent a critical point as for the thermoelectric voltage which
can disturb the result. Thermoelectric voltages are large for
combinations of Cu-Cu(oxide), Cu-Ni and Cu-solder(Sn/Pb) and
amount to $\approx$1000~$\mu$V/$^\circ$C, 22~$\mu$V/$^\circ$C and
1--3$\mu$V/$^\circ$C, respectively. On other side, the
thermoelectric voltage for Cu-Au is only of 0.2~$\mu$V/$^\circ$C.
For this reason the nickelled brass connectors of the divider were
replaced by gold plated copper connectors. The connecting twisted
leads were produced from copper wire of about 50~cm and
equipped with copper gold plated spade terminations soldered by
special Sn/Cd alloys (thermoelectric voltage for
Cu-alloys(Sn/Cd) is low and amounts to 0.8~$\mu$V/$^\circ$C).
The FLUKE voltmeter terminals are not affected by undesirable
thermoelectrical voltage as they are
made from tellurized copper which is resistant to oxidation.
Before starting HV divider ratio measurement, several
hours have to pass to minimize the temperature gradients at the
divider and the voltmeter.

c) \\
The spectrometer vacuum pumping system based on an old titanium getter
pump with a double stage rotary pump (5~m$^{3}$/h) as a backing
pump reached a vacuum level of about
2$\times10^{-7}$~mbar. Replacing the titanium pump by a 210~l/s
turbomolecular pump improved the vacuum to 2$\times10^{-8}$~mbar.

d) \\
The original vacuum gauges controlling the spectrometer safety
system were cooled by an open water system with a consumption of
1.2 m$^{3}$ of water per day. Practically continuous operation
of the spectrometer is expected during the solid
$^{83}$Rb/$^{83\rm{m}}$Kr source development. In order to
avoid large costs for water a closed water circuit equipped with
pump and small refrigerator was introduced. The mean temperature of
cooling water is set to 11 $^\circ$C.

e) \\
In the spectrometer volume, conversion electrons are
produced by decay of $^{83\rm{m}}$Kr released from the source.
These electrons scatter on spectrometer material and produce
secondary electrons. Some of them are detected by the channeltron
as a background. Applying a low voltage of -29~V to the channeltron input
funnel the background in the electron spectra was reduced by factor of 4,
whereas the line intensity at pass energy $E_{an}=260$~eV was
reduced only by factor of 1.6. A similar procedure was used in
Dubna at spectrometer ESA50 \cite{Ino86}.

f) \\
The spectrometer control program was equipped with configuration input
files to be able to run measurements with any regime using any
combinations of all available
voltmeters, HV dividers and HV power supplies. If necessary, the measurement
of temperature using the SOLARTRON temperature gauge can be also
selected in the configuration file. Because the integration time
for FLUKE voltmeters with a resolution 7.5 digits is  1.8~sec,
the time devoted to count in the counter had to be
prolonged to 2~sec. The structure of the spectrometer output data file was also
changed, currently in each sweep and for each point the actual values
of $E_{lens}$, $E_{an}$, counts in 2~sec and ambient temperature are recorded.

g) \\
The old computer code for the spectrum sorting from output file
occurred to be insufficient for the development of the radioactive
source with reproducibility of the conversion electron energy at
sub-electronvolt level. In the old code the appropriate energy
window $\Delta E$ was chosen and all counts corresponding to the
energy bin $b_{i} \equiv  [E_{i}-(\Delta E / 2), E_{i} + (\Delta E
/ 2) ]$ were summed up together. The value of the energy $E_{i}$,
bin energy, was ascribed
to this sum of the counts. Usually, $E_{1}=E_{begin}$
and $\Delta E=E_{step}$; obviously, $E_{i+1}=E_{i}+\Delta E$,
\cite{Dra95}.  Unfortunately, due to the HV supply imperfections
for the steps of several tenths of
volts the sorted spectra exhibited count irregularities and
the determined line position was unreliable. Therefore, three new codes
were written. The first one does the overall diagnostic for the
format correctness of the output file and sorts four dependencies
from which the quality of the measurement can be estimated:
\begin{itemize}
\item{D1: frequency distribution of values of the energies for the
      measured points (i.e.
      $eU_{lens}+(1/c_{an})\times|U_{an}|$ value) using bins (usually
      the bin width is taken as 0.05~eV) which are successively set to
      cover without spaces the range ($E_{begin}-2, E_{end}+2$)~eV  }
\item{D2: frequency distribution of values of the steps (i.e. $E_{step}$,
      the difference between the neighboring energy points in measured range)
      using bins (usually 0.01~eV) which are successively set to cover
      without spaces the range (0, $E_{step}+2$)~eV  }
\item{D3: dependence of the $eU_{lens}+(1/c_{an})\times|U_{an}|$  values for
      particular energy point of the measured range (usually the points 1, 2 and 3
      are chosen) on the sweep number}
\item{D4: dependence of ($(1/c_{an})\times |U_{an}|$) values
      and temperature --- both quantities averaged for points
      in frame of given sweep --- on the sweep number.}
\end{itemize}
In Figs.~\ref{D1}---\ref{D4} examples of dependencies D1-D4 are
displayed for three day measurement with the
$^{83}$Rb/$^{83\rm{m}}$Kr solid source: line L$_{1}$-9.4,
$E_{begin}=7467$~eV,  $E_{end}=7496$~eV, $E_{step}=0.5$~eV,
altogether 2500 sweeps.
It is visible from Figs.~\ref{D1} and \ref{D2}  that in a real
measurement, the actual step was set to 0.59~eV although in
control program input data a value of 0.5~eV was required, which
in terms gives 50 energy points in the range instead of 59. This
is a result of non ideal transformation of the digital signal into
analog one in the HV power supply. In Fig.~\ref{D3} some
instability at a level of 200~meV is visible at the beginning of the
measurement and in the second half of the measurement, starting at
sweep number $\approx$~1500. This effect was connected with lens
HV supply as $U_{an}$ was stable within 8~mV, see Fig.~\ref{D4}.
In Fig.~\ref{D4}, one can further clearly see the day and night
temperature variation and corresponding variation of cylinder
electrode voltage.

The second computer code is designated for sorting of the electron
spectrum itself. It works similarly as the previous code
but this time the bin structure for spectrum sorting is taken from frequency
spectrum in Fig.~\ref{D1}. Ideally, $\delta$-functions would be
seen; instead peaks of finite width and variable
amplitude are visible. For correct sorting of the energy spectrum,
the centroids of these peaks were chosen to serve as the bin energy.
In such a way the bin widths for different points are different.
The width for a particular bin is obtained as a sum of the half
distance between the bin centroid and adjacent bin
centroids. In this case, the count irregularities (present in
spectra sorted according to the old approach) disappeared.

The third code is also designated for sorting of the electron
spectrum. This time a new, direct, no-bining method is applied.
The conversion line energies determined using the spectra sorted
by this code are within statistical error the same as those
determined by the second code. But
generically, the new method is better as it takes into account
all individual values of voltages measured for each point in each
sweep. The full description of the no-bining method can be found
in \cite{Rys08}.

\subsection{Electrostatic retarding MAC-E-Filter spectrometer
at The Institute of Physics, University of Mainz}
 \label{245mac-e-filter}
The electron spectrometer of Mainz university is of the
MAC-E-Filter type, see \cite{Pic92}. The device was used for
the neutrino mass experiment with tritium \cite{Kra05}.
It consisted of quenched condensed tritium source (QCTS),
an electron transport section (ETS), MAC-E-Filter spectrometer with
vacuum vessel, retarding electrode and two superconducting pinch
magnets and detector with superconducting magnet. Additionally, the device was
equipped with central air coil controlling the low magnetic
field in center of the spectrometer vacuum vessel and two mutually perpendicular
earth compensation coils for the elimination of the influence of
the earth magnetic field. After bake up a water free vacuum with final
pressure of about $5 \times 10^{-10}$~mbar was achieved.

\begin{figure}[h]
 \begin{center}
 \includegraphics*[bb= 13 208 579 631 pt, clip=true,
  width=11cm,totalheight=8cm]{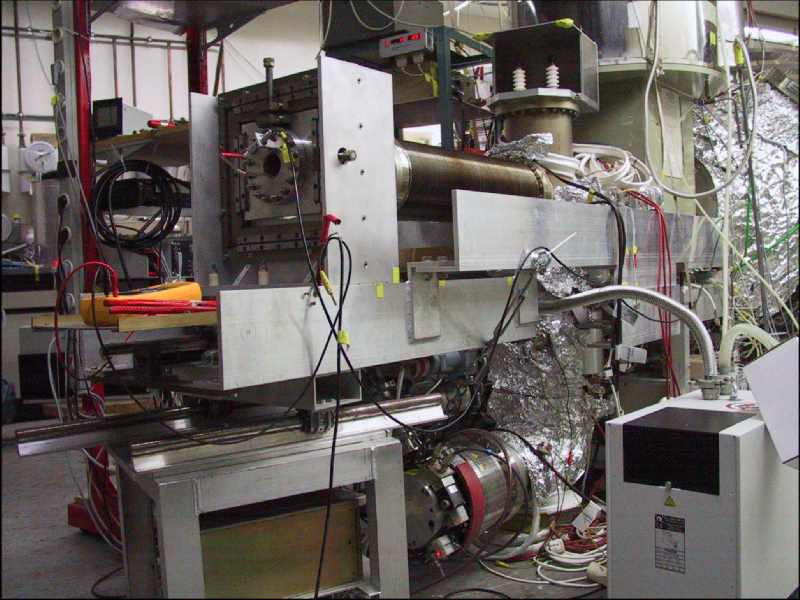}
 \caption[fig21]{Source part of the Mainz MAC-E-F electron retarding
                 spectrometer. On the right side the source pinch magnet and part
                 of spectrometer vacuum vessel are also seen.}
 \label{MAINZ}
 \end{center}
\end{figure}

\begin{figure}[th]
 \begin{center}
 \includegraphics*[bb= 20 346 404 597 pt, clip=true,
  width=10cm,totalheight=6cm]{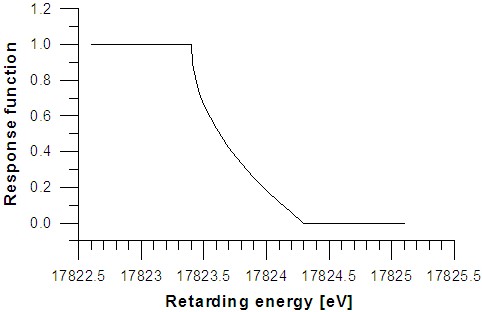}
 \caption[fig22]{Response function of the Mainz MAC-E-Filter
  spectrometer for the instrumental resolution of 0.9~eV and
  K-32 energy $E_{0} = 17824.3$~eV. The spectrometer work function
  and change of binding energy in solid source are in this
  example omitted.}
 \label{resfun}
 \end{center}
\end{figure}

At present the spetrometer is used for systematic studies
necessary for the KATRIN project. A HV supply for 35~kV from
F.u.G. ELECTRONIC is used for the retarding electrode. The
high voltage is measured by means of the newly developed KATRIN HV
divider \cite{Thu07} and voltmeter FLUKE 8508A. The divider ratio
and the divider ratio drift amount to $\approx$~1972.4 and +0.15~ppm/week,
respectively. The FLUKE voltmeter is daily calibrated
at --10 V using FLUKE 732A DC Reference Standard. The typical value of drift
of the FLUKE 732A reference is 0.06 ppm/week. From half a year
measurements follows that the FLUKE voltmeter has at voltage
--10~V a drift of --0.22 ppm/week. For measurements with solid
$^{83}$Rb/$^{83\rm{m}}$Kr sources, the ETS and QCTS were replaced
by a new system allowing installation of the source. Moreover, the
source being in vacuum can be aligned in X,Y and Z directions
relative to the spectrometer. The Z direction is defined by the
spectrometer horizontal axis along which are the magnetic guiding and
electric retarding fields
axially symmetric. The source part with the X-Y-Z source alignment
mechanism and the 185~mm diameter compensation bellows are seen in
Fig.~\ref{MAINZ}.
The detector magnet was removed, too. The detector had to be moved
closer to the pinch magnet in order to conserve suitable detector
count rate. The spectrometer measures integral spectra in an
energy region of 7--35~keV. The lower limit is given by electronic
noise of the detector. The upper limit is
dictated by the maximal possible voltage which can be applied to
the retarding electrode. The instrument enables us to measure not
only the K-32 conversion electrons but also other conversion
electron lines observed in $^{83\rm{m}}$Kr decay. For the 17.8~keV
energy of K-32 electrons, usually instrumental energy resolutions
of 2.0 or 0.9~eV are used. These values are achieved by applying
the magnetic field (6.014~T) in the pinch magnets and by setting a
low magnetic field in central retarding plane. The response
function $T(E_{0},E)$ for the electrons with energy $E_{0}$ and
instrumental resolution $\Delta E$ is described by the following
relations: \\
\\
\\
$T(E_{0},E) = 1$ \,\, if \,\, $E \leq E_{0}- \Delta E$, \\
\\
$T(E_{0},E) =\frac{1 - \sqrt{ 1-\frac{(E_0-E)}{\Delta E}
                        \times \frac{B_S}{B_P}  }        }
                 {1 - \sqrt{ 1-\frac{B_S}{B_P}  }        }$
            \,\, if \,\, $E_{0}- \Delta E< E < E_{0}$, \hspace{3cm} (3)\\
\\
$T(E_{0},E) = 0$  \,\, if \,\, $E \geq E_{0}$, \\
\\
where \\
\begin{itemize}
\item{$E = qU$ is the retarding energy for the retarding voltage $U$,}
\item{$E_{0}$ represents the energy of monoenergetic electrons,}
\item{$B_S$ is the magnetic field at the electron source,}
\item{$B_P$ corresponds to the maximum magnetic field at the pinch magnet.}
\end{itemize}
This response function definition holds under an assumption
that the electrons are leaving the source isotropically.
In Fig.~\ref{resfun} an example of response function for
the energy of K-32 conversion electrons is shown.
The spectrometer transmission depends on distance between the
source and the center of the pinch magnet on the source side. For
example, for distances of 25, 20 and 15~cm, the transmission
amounts to 3.3, 10.0 and 23.5~\% of $4\pi$, respectively. The old,
partly damaged, semiconductor silicon chip was replaced by a new
detector, a pin diode from HAMMAMATSU. The diode with an area
9~mm $\times$ 9~mm and max. HV --150~V is cooled to --90$^\circ$C.
In front of the detector a mask was installed with diameter 9~mm.
The energy resolution for 17.8~keV electrons is 2.1~keV.
and the efficiency for registration of electrons is of 80~\%.
For processing of the detector signals a standard electronic
chain with the cooled FET preamplifier, spectroscopic amplifier
and a 8192 channel ADC is used. A histogrammig memory was also
applied in order to reduce the dead time of the data acquisition
system. The correction for the dead time is provided by means of
pulser signals applied to the preamplifier pulser input. During
the measurement a computer is controlling the HV supply, voltmeter
and data acquisition system.

\section{Measurements and results}
\label{4measres}
\subsection{Sources preparations and measurements in \v Re\v z}
\label{41rez}
The development of the source of conversion electrons with an energy
stability in the ppm level for at least 2 months is a
difficult and time consuming process. Obviously, only to find out
this source quality needs tests with the source at the electron
spectrometer for 2-3 months. Because our first measurements showed
unsatisfactory reproducibility and stability of the energy of the
L$_{1}$-9.4 line a large effort was devoted to study this effect by
improvements of the spectrometer (see Sect.~\ref{34esa12}).
Attention was also payed to increase the source activity in
order to decrease statistical errors.
Finally, different modes of source preparation were tested.

During the period 2006-2008, 6 irradiations with krypton gas
targets were performed at NPI. From the activity produced, 21 solid
$^{83}$Rb/$^{83\rm{m}}$Kr sources were prepared by vacuum
evaporation and measured at gamma and electron spectrometers. The
list of sources according to the irradiation and source production
dates is shown in Table~\ref{tab2}. Further properties as a boat
activity, substrate type, boat---substrate distance in evaporation
device, efficiency of evaporation, source activity, retention of
$^{83\rm{m}}$Kr in the source and reduced amplitude of L$_{1}$-9.4
line $A_r$ are also indicated.
\begin{table}[ht]
\caption{Solid $^{83}$Rb/$^{83\rm{m}}$Kr sources in period
2006-2008.}
\begin{tabular*}{6.5in}[b]{ccccccccrc}
\hline
Irrad. & Source No. & Prod. & Boat act. & Substr. & Dist. & Eff.ev. & Activ. & Ret. & A$_r$ \\
&      &       &  [MBq] &  &[mm]  &[\%]  &  [MBq]  &  [\%]  &            \\
\hline
02.11.05 &   8$^{a1}$ & 29.11.05   &89 & Al  &  14 &  5 &  4.6 &   70 &   0.56 \\
         &   9$^b$    & 17.02.06   &34 & Al  &  14 &  6 &  2.1 &   25 &   0    \\
         &  10$^b$    & 22.02.06   &23 & Al  &  14 &  3 &  0.6 &   53 &   0    \\
10.05.06 &  11        &  29.05.06  &140& Al  &  14 &  2 &  2.4 &   18 &   0.63 \\
         &  12$^b$    &  27.06.06  &-  &  Al &  14 &  - &  5.8 &  100 &   0.03 \\
29.11.06 &  13        &  30.01.07  &-  &  Al &  11 &  - &  0.5 &    7 &   -    \\
         &  14$^{a2}$ & 30.01.07   &-  & Al  &  10 &  - &  0.4 &    5 &   -    \\
         &  15        &  06.02.07  &66 &  Al &  11 & 10 &  6.6 &   12 &   1.01 \\
13.06.07 &  16        &  02.07.07  &11 & HOPG&  13 &  9 &  0.9 &   13 &   0.96 \\
         &  17        &  03.07.07  &93 &  Al &  11 & 13 & 12.0 &   11 &   0.64 \\
         &  18        &  04.07.07  &10 &  C  &   7 & 59 &  6.1 &   23 &   0.65 \\
30.11.07 &  19        &  27.03.08  &25 &  C  &  12 & 11 &  2.7 &    3 &   0.95 \\
         &  20$^c$    & 27.03.08   &18 & C   &   7 & 28 &  4.9 &   30 &   0.41 \\
         &  21        &  27.03.08  &8  &  C  &   5 & 38 &  3.0 &    7 &   0.97 \\
09.04.08 &  22$^d$    & 23.04.08   &10 & C   &   1 & 20 &  2.0 &   33 &   0.52 \\
         &  23$^d$    & 29.04.08   &10 & C   &   3 & 46 &  4.7 &   60 &   0.27 \\
         &  24$^d$    & 29.04.08   &11 & C   &   3 & 30 &  3.2 &   49 &   0.26 \\
         &  25$^d$    & 07.05.08   &10 & C   &   3 & 32 &  3.2 &   87 &   -    \\
         &  26$^d$    & 07.05.08   &9  & C   &   3 & 55 &  5.1 &   63 &   0.26 \\
         &  27$^d$    & 16.06.08   &15 & C   &   6 & 24 &  3.8 &   73 &   0.21 \\
         &  28        &  28.07.08  &18 &  C  &   5 & 27 &  5.0 &   19 &   0.53 \\
         &  29        &  28.07.08  &19 &  C  &   5 & 26 &  4.9 &   11 &   0.62 \\
\hline
\multicolumn{10}{l}{$^{a1,a2}$ The $^{83}$Rb water solution with LOCTITE glue added,
  for details see text.} \\
 \multicolumn{10}{l}{$^b$ The $^{83}$Rb for deposition into the boat  obtained from
  the $^{83}$Rb contaminated Al protection foil.} \\
\multicolumn{10}{l}{ $^c$ The non-carrier free deposit in the boat - to the $^{83}$Rb
  a non-active rubidium was added.} \\
\multicolumn{10}{l}{ $^d$ $^{83}$Rb purified by aerosol filter was deposited
  into the boat, for details see text.} \\
 \end{tabular*}
\label{tab2}
\end{table}
The amplitude $A_r$ in the table is given by the relation
\begin{equation}
  A_{r} = 32 \times H / A / (23+77/100 \times R),
\end{equation}
where
\begin{itemize}
\item{32 is scaling constant,}
\item{$H$ represents the measured
amplitude of the L$_{1}$-9.4 conversion electron line in $s^{-1}$,}
\item{$A$ is the mean activity of the source in MBq for the period
of measurement,}
\item{$R$ is retention
of $^{83\rm{m}}$Kr in the solid source in \%.}
\end{itemize}
The reduced amplitude enabled us to compare sources with different
activities and retentions. The numbers in denominator 23 and 77
are the electron capture and $^{83\rm{m}}$Kr isomer feeding
intensity of the level at 9.4 keV energy, respectively. These
numbers are deduced from nuclear decay data, see also
Fig.~\ref{schemea}. The different values of A$_r$ indicate the
different amount of zero-energy loss electrons and consequently a
different contamination present in the sources. The initial
conditions for the particular source production are indicated in
the columns Boat activity, Substrate and Distance. For some
sources additional conditions, marked by indexes $^{a,b,c,d}$ in
the second column, are explicitly described below the table.
The last 4 columns represent the source property itself.

\begin{figure}[ht]
 \begin{center}
 \includegraphics*[bb= 19 360 396 596 pt, clip=true,
  width=10cm,totalheight=6.0cm]{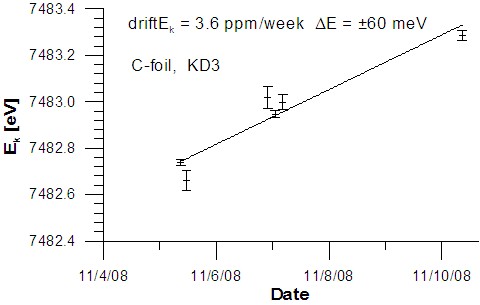}
 \caption[fig26]{Dependence of the L$_1$-9.4 line position on the date
  of measurement for source No.~20.}
 \label{pos20}
 \end{center}
\end{figure}

The Table~\ref{tab2} reflects four dedicated tests:

a) \\
At the production of the source No.~8 accidently the rubidium
water solution  was polluted by LOCTITE glue used in the
construction of the first version of the gas target. Nevertheless,
the source quality was good - the retention and reduced amplitude
were high. In order to repeat this favorable case the LOCTITE glue
was added when the source No.~14 was produced. Unfortunately, the
expected high retention was not achieved.

b) \\
During the vacuum
evaporation a large portion of rubidium activity is deposited on
the protection foil used in evaporation chamber. Drops of
distilled water were applied on this foil and later put away and
purified. The procedure for this purification was the same
as those used for the water rubidium solution obtained
from gas target. Sources prepared in this way No.~9, 10, 12 indicated
large krypton retention, but the amplitude of the zero-energy loss
line was practically zero.
\\

c) \\
The sources were mostly
carrier free with the calculated mean thickness of about 0.1--0.4
monolayers of radioactive rubidium. The solid source No.~20 was
prepared with addition of nonactive rubidium carrier in amount of
10 monolayers in order to define better the environment of the
$^{83}$Rb atoms.
Into the water solution with 18 MBq of purified $^{83}$Rb  16~$\mu$g
of RbNO$_3$ (purity 99.99) were added. Finally, the solid source
contained 2~\% of the radioactive rubidium and 98~\% of the stable
rubidium.
No marked result was achieved, the krypton
retention was higher of a 30~\% and the reduced amplitude was lower
but still acceptable $A_r=0.41$. The L$_{1}$-9.4 line position
exhibits a drift of 3.6~ppm/week, see Fig.~\ref{pos20}.

d) \\
The more simple and faster purification of the rubidium water solution
by means of aerosol filter instead of chromatography column was
tested. The sources No. 22--27 document the results. The Kr
retention was high. The reduced amplitude was
acceptable of $\approx 0.5$ when the source was fresh but after
one month it became too low of $\approx0.26$. The energy of
the L$_{1}$-9.4 electron line exhibited fully unacceptable large
negative drifts of about --40 ppm/week. For this reason the
aerosol filter technique was rejected from further development.

At an early stage the sources were evaporated onto a Al substrate, similarly
as it is described in \cite{Kov92} and \cite{Kov93}.
Because the energy of L$_{1}$-9.4 line exhibited a non zero drift
 and in order to avoid possible
problem with insulating aluminium oxide inherently present on Al foils
the next sources were prepared on carbon substrates. Nevertheless,
a substantial reduction of the drift or a change in krypton retention
with carbon substrate was not observed.

For the first evaporations the boat-substrate distance was kept
large, of about 14~mm. The corresponding efficiency of the vacuum
evaporation process was only 2--11~\%. At the closer distance
the efficiency of the vacuum evaporation process was expected to
be even lower. Namely, the $^{83}$Rb deposited originally on the
substrate could evaporate again due to the heat from the boat. However,
the tests with closer
boat---substrate distances were successful and much higher efficiencies of
several tens of percent, were achieved, see Table~\ref{tab2}. The
dependence between efficiency and the boat---substrate distance is
not always smooth. This is very probably caused by the effect of
the boat deformation during the boat heating described in
Sec.~\ref{32evap}. Due to this effect the boat-substrate distance can
increase or decrease up to 2~mm. The larger value of
evaporation efficiency observed for the sources No.~18 and 21 is
probably caused by a particular distribution of rubidium
deposit in the boat.

\begin{figure}[ht]
 \begin{center}
 \includegraphics*[bb= 19 360 396 596 pt, clip=true,
  width=10cm,totalheight=6.0cm]{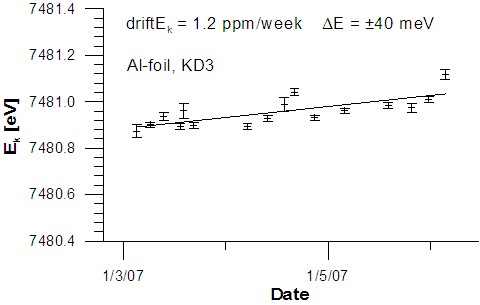}
 \caption[fig23]{Dependence of the L$_1$-9.4 line position on the date
  of the measurement for source No.~15.}
 \label{pos15}
 \end{center}
\end{figure}

\begin{figure}[ht]
 \begin{center}
 \includegraphics*[bb= 19 360 396 596 pt, clip=true,
  width=10cm,totalheight=6.0cm]{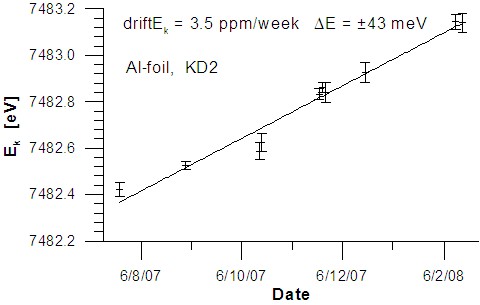}
 \caption[fig24]{Dependence of the L$_1$-9.4 line position on the date
  of measurement for source No.~17.}
 \label{pos17}
 \end{center}
\end{figure}

\begin{figure}[ht]
 \begin{center}
 \includegraphics*[bb= 19 360 396 596 pt, clip=true,
  width=10cm,totalheight=6.0cm]{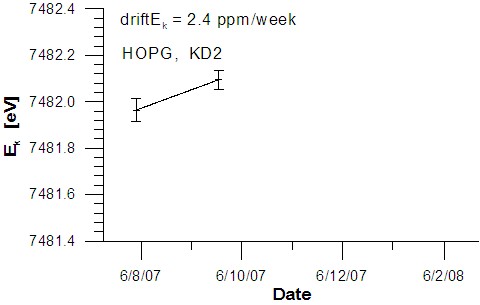}
 \caption[fig25]{Dependence of the L$_1$-9.4 line position on the date
  of measurement for source No.~16.}
 \label{pos16}
 \end{center}
\end{figure}

\begin{figure}[ht]
 \begin{center}
 \includegraphics*[bb= 19 360 402 596 pt, clip=true,
  width=10cm,totalheight=6.0cm]{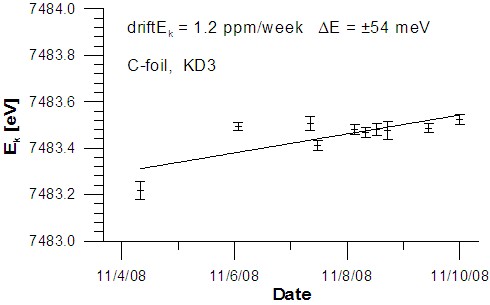}
 \caption[fig27]{Dependence of the L$_1$-9.4 line position on the date
  of measurement for source No.~21.}
 \label{pos21}
 \end{center}
\end{figure}

In Figs.~\ref{pos15}--\ref{pos21} the dependences of the L$_1$-9.4
line position on the date of the measurement for different sources are demonstrated.
In the figures the type of source substrate and the divider applied
(KD2 or KD3) are also shown.
For sources and periods of measurements:
\begin{itemize}
\item{No.~15 \,\, 5. 3. 07 - 31. 5. 07,}
\item{No.~17 \,\, 6. 10. 07 - 7. 12. 07,}
\item{No.~21 \,\, 15. 08. 08 - 30. 11. 08}
\end{itemize}
a particular source was continuously in the spectrometer, i.e. at
vacuum condition, occasionally at retarding voltage of
$\approx$~7200.~V. The other points in the figures were obtained
immediately after installation of the given source into the
spectrometer being previously kept at air. The errors of the
energies are only the statistical ones. In all cases the energy of
the conversion line is increasing with time of the measurement and
can be described by linear dependence. The corresponding linear
fit is indicated in the figures together with the value of line
drift in ppm/week and the mean scatter $ \Delta E$ of the
individual measurement relative to the linear fit in meV.

After correction for the drift of KD2 or KD3 divider the drifts of
the L$_{1}$-9.4 line positions are in the range 0.6--2.4 ppm/week.
The correction to the voltmeter drift is omitted at this stage of
analysis because it is relatively small. Besides, the different
sources exhibit different starting value of energy, i.e. the
energy measured immediately when the source was installed into the
spectrometer. It must be stressed that sources were prepared in
different way (e.g. different distance boat-substrate or the
activity of the deposit in a boat) originally with the aim to test
the modes of the source preparation. Nevertheless, it seems that
the properties of the sources are sensitive to the mode of their
preparation. This effect can be observed also in the cases where
the initial conditions for the source production are quite close
together, see e.g. sources 28 and 29. In spite of this the krypton
retentions are different, they amount of 19 and 11~\%,
respectively. Moreover, the first short measurement of sources at
ESA12 (just before sending the sources for the measurements at
Mainz spectrometer) gave the values of energies for the L${1}$-9.4
line which differed by 0.7~eV. Strictly speaking there was a
slight difference during production of these sources. When the
rubidium water solution for the 29 source boat deposit was
prepared an amount of distilled water was added into the eppendorf
vessel with rubidium activity in order to wash out all
last rubidium activity in the vessel.

On the other side, the
mean scatter of conversion line energies along the fitted drift
line is reaching only $\pm60$~meV for a period of 3 months or more.

\subsection{Measurements in Mainz}
\label{42mainz} The measurements using the sources No.~11 and 13
with activities of several hundreds of kBq were performed in 2007.
The sources were attached to the cold finger of the
existing condensed krypton source setup, consequently the source
centering in X-Y plane was very limited, however measurements
with cooled sources were also possible. Due to the elimination of
the detector magnet and without possibility to move the detector
closer to the pinch magnet the detector count rate was reduced and
the spectra suffered by lower statistics than
expected --- the count rate for the K-32 conversion electrons was
only 30~$s^{-1}$. By cooling the source to
$\approx$~119~K (the boiling point of krypton) the detector rate
increased up to 400 $s^{-1}$. Evidently, the krypton retention increased
to 100~\%. In this case the statistical error for the energy of
K-32 electrons approached 10~meV after 12~hours of
measurement. Unfortunately, the long term operation with cooled
sources is not suitable. Condensation of the rest gas
on the source changes its properties, including the energy of
emitted electrons.

\begin{figure}[ht]
 \begin{center}
 \includegraphics*[bb= 20 349 388 595 pt, clip=true,
  width=10cm,totalheight=6cm]{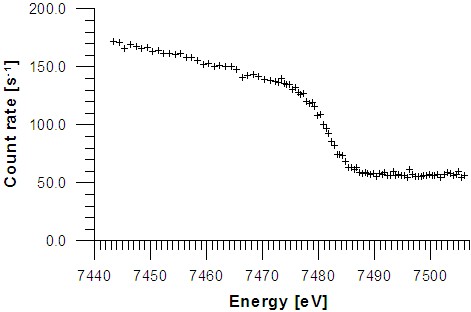}
 \caption[fig28]{Spectrum of L$_{1}$-9.4 electrons measured with
 instrumental resolution set to of 0.4~eV.}
 \label{L1}
 \end{center}
\end{figure}

\begin{figure}[ht]
 \begin{center}
 \includegraphics*[bb= 20 351 393 595 pt, clip=true,
  width=10cm,totalheight=6cm]{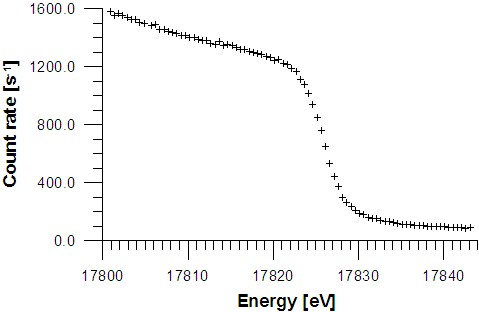}
 \caption[fig29]{Spectrum of K-32 electrons measured with instrumental
  resolution set to 0.9~eV.}
 \label{K}
 \end{center}
\end{figure}

\begin{figure}[ht]
 \begin{center}
 \includegraphics*[bb= 21 352 409 595 pt, clip=true,
  width=10cm,totalheight=6cm]{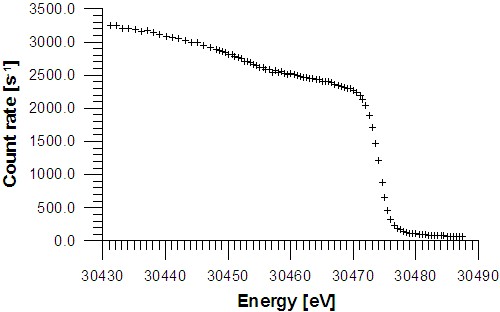}
 \caption[fig30]{Spectrum of L$_{3}$-32 electrons measured with instrumental
  resolution set to 1.6~eV.}
 \label{L3}
 \end{center}
\end{figure}

The next measurement in Mainz was started in August 2008 and will
be finished during the first quarter of 2009. For this experiment
vacuum evaporated rubidium sources No.~28 and 29 with activities
of 5~MBq each were prepared, see Table~\ref{tab2}. Now the sources
could be aligned by means of the X-Y-Z system. Till now only the
results of the source No.~28 are available. After installation of
the source into the spectrometer, the measurements of spectra of
different conversion electrons were performed for a period of 80
days. Using the same spectrometer distance source-magnet center as
in 2007 year a substantially larger count rate of about
1200~s$^{-1}$ for K-32 electrons was achieved. A statistical error
of 10~meV could be obtained within 3 hours of measurement. The
time of one individual measurement was 1-3 hours. In
Figs.~\ref{L1}--\ref{L3} typical integral spectra of conversion
electrons L$_{1}$-9.4, K-32 and L$_{3}$-32 are displayed.
The time of the measurement at a particular voltage was 40
seconds. The part of the spectrum where the step occurs is
important for the precision of the line position determination.
This region was measured with smaller voltage step in order to
accumulate larger statistics. The spectrum in Fig.~\ref{L1}
exhibits low statistics although the L$_{1}$-9.4 line is the
strongest in $^{83}$Rb decay. It was not possible to extract
the full count rate due to the overlap of the line with
the detector noise .

\begin{figure}[ht]
 \begin{center}
 \includegraphics*[bb= 18 356 405 595 pt, clip=true,
  width=10cm,totalheight=6cm]{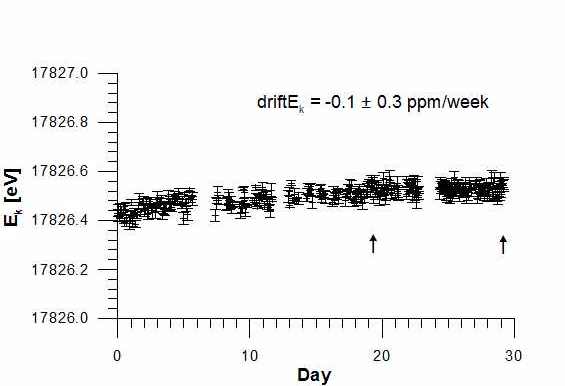}
 \caption[fig31]{Dependence of the K-32 line position on time
  within the days 1--30; before 0.9~eV jump in K-32 line energy.}
 \label{mainz1}
 \end{center}
\end{figure}

\begin{figure}[ht]
 \begin{center}
 \includegraphics*[bb= 18 356 405 595 pt, clip=true,
  width=10cm,totalheight=6cm]{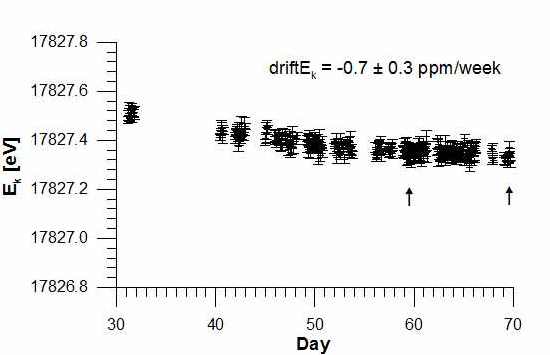}
 \caption[fig32]{Dependence of the K-32 line position on time
  within the days 30--70; after 0.9~eV jump in energy of K-32.}
 \label{mainz2}
 \end{center}
\end{figure}

\begin{figure}[ht]
 \begin{center}
 \includegraphics*[bb= 18 356 405 595 pt, clip=true,
  width=10cm,totalheight=6cm]{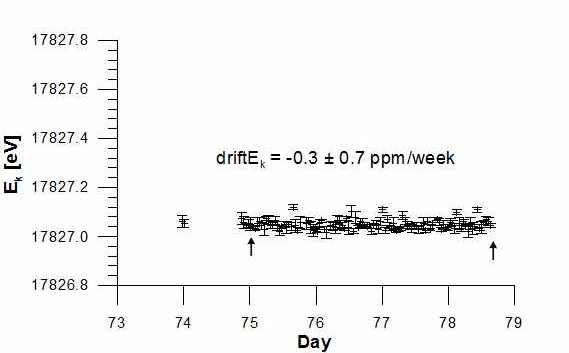}
 \caption[fig33]{Dependence of the K-32 line position on time
  within the days 70--82; after baking up of the source.}
 \label{mainz3}
 \end{center}
\end{figure}

In the experiment the main attention was paid to the stability of
the energy of the K-32 conversion line. The 80 days of measurement
can be divided into 3 periods. The time dependence of the K-32 line
position for these periods is shown in
Figs.~\ref{mainz1}--\ref{mainz3}.
The values of the energies $E_k$ are corrected for the drift of
the voltmeter. On the other side, the drift of the dividing ratio
(+0.15 ppm/week) is not taken into account. The region of days where
the K-32 line position was most stable, is indicated in each figure
by arrows --- the corresponding value of the drift is added. The
first period includes the measurements started two days after
source installation and finished on the 30th day where the jump of
0.9~eV in energy of K-32 line appeared. This jump was accompanied
by the failure of temperature regulation of the HV divider. Due to
the failure the temperature inside the divider increased. At the same
time when the HV divider failed one of the two supplies for
the earth field compensation coil failed also. So the magnetic field
lines from spectrometer filter electrode to the source may appear.
This could guide electrons and/or negative ions (e.g. H$^{-}$) to the
source. Since they are accelerated with the filter voltage, they
could do some changes to the source surface.
In the first period the K-32 energy is raising with decreasing slope ---
an effect of saturation is visible, see
Fig.~\ref{mainz1}. After divider and supply repair the measurement continued,
however the jump in energy of the K-32 line remained. The same
jump was observed also for the L$_{1}$-9.4, L$_{3}$-32 and
N$_{2,3}$-32 lines. In spite of large efforts devoted to this event
the reason for such a large and sudden jump in energy remains
unexplained. The influence of the HV supply, HV divider or
voltmeter was excluded by the dedicated tests. The change of the
source surface by negative ions and the sudden change
in the spectrometer work function or in source property itself seems to
be the only reasonable explanation for the observed jump effect.
The measurements of the K-32 spectra after the jump in energy show again
saturation, this time the energy is decreasing, see
Fig.~\ref{mainz2}.

After 70th day the source was heated up to
70$^\circ$~C for 10 hours in order to test the heating procedure
itself (the heating is intended to be used for the next source
No.~29 in order to speed up again the expected saturation effect). For
heating the source was moved back and
isolated by a vacuum valve from the spectrometer. After heating,
when the source temperature was again at room temperature, the
measurements of the lines K-32, L$_{1}$-9.4, L$_{3}$-32 and
N$_{2,3}$-32 were performed. The energies for all lines were
increased by 25~meV. The amplitude of the K-32 electron line was
reduced by only 2~\%. Altogether, the heating procedure is
applicable for the solid $^{83}$Rb/$^{83\rm{m}}$Kr source. For the
next measurements source and detector were moved closer to the
source and detector magnets as the source activity was reduced due
to the radioactive decay. The detector rate for the K-32 lines
increased by factor 3. The result of 5 days measurement is
displayed in the Fig.~\ref{mainz3}. Due to the higher statistics
the band of energy values $E_{k}$ is slightly more narrow in
comparison with measurements in the previous period. But the values
$E_{k}$ are reduced by 0.36~eV. This decrease is a result of the
source and detector geometry change. It indicates the still
insufficient alignment of the source and detector.

 The drifts obtained for the K-32 energy saturation for
the first and third period are compatible with the stability
requirement $\pm$60~mV in 2~months.

\section{Conclusion and Outlook}
\label{5conout} The experience gained during the development of the
solid $^{83}$Rb/$^{83\rm{m}}$Kr sources showed that all future
sources must be produced as reproducibly as possible. This
concerns also the activity which itself must be about 10~MBq.
Sources with such activity and moderate $^{83\rm{m}}$Kr retention
of 10--20~\% can be used at the monitoring spectrometer for 4--6
months. The future measurements at ESA12 spectrometer will be mainly
for long term stability assessment ($\approx$~2 months), keeping the source
all the time in vacuum. The technique of source heating
could be introduced.

The results of the measurement at Mainz
are promising. Unlike to the results at \v Re\v z, an
effect of saturation of the K32 line energy value with time was
observed. Of course, the next measurement with the source
No.~29 will be very important. The present result of the Mainz
measurements with the source No.~28, indicates that such a source
has a perspective to be used for monitoring at least for 2~months,
a typical time of the KATRIN individual experimental run.
Especially, the observation of the very small drifts 0.1(3) and
-0.3(7)~ppm/week of the K32 line in periods were the source-spectrometer
system was stable is encouraging.

\section{Acknowledgment}
\label{6ackn}
We thank to M.~Fi\v{s}er and J.~Stanislav for the design and production
of the krypton gas target. We are also obliged to P. Han\v{c} for
handling of the krypton gas target.
The authors wish to thank M.~Beck and H.W.~Ortjohann
for the help with the design and production of equipment for the
source installation at Mainz spectrometer. We wish to express our
gratitude to B.~Ostrick for fruitful discussions during
measurements in Mainz. We are indebted to
T.~Th\"{u}mmler for valuable advices concerning the KATRIN HV divider
and precise DC voltage measurement.
We gratefully acknowledge valuable communications with E.W.~Otten.

This work was performed under the auspices of the Academy of
Science of the Czech Republic under contract ASCR IRP AV0Z10480505.
Further support by the Grant Agency of the
Czech Republic under contract No. 202/06/0002 and by the M\v{S}MT
under contracts LA318 and LC07050 is highly appreciated.
Two of us (D.~V\'{e}nos and A.~Koval\'{i}k) were supported by
the Deutsche Forschungsgemeinschaft. We acknowledge University of
Mainz and the hospitality of the Institute of Physics for performing
the measurements at the Mainz spectrometer. The Mainz and M\"{u}nster
groups were supported by grants of BMBF.

\end{document}